\documentclass[prl,aps,twocolumn,superscriptaddress,showpacs,floatfix,letterpaper]{revtex4}
\usepackage{subfigure}
\usepackage[dvips]{graphicx}
\usepackage{amsmath}
\usepackage{epsfig}

\begin{document}

\def\q{{\bf q}}
\def\k{{\bf k}}
\def\Q{{\bf Q}}
\def\r{{\bf r}}

\title{Detecting D-Wave Pairing and Collective Modes in Fermionic Condensates with Bragg Scattering }

\author{G.~R.~Boyd}
\affiliation{Condensed Matter Theory Center University of Maryland, College Park, MD 20742-4111, U.S.A.}
\author{V.~Galitski}
\author{V.~M.~Yakovenko}
\affiliation{Condensed Matter Theory Center University of Maryland, College Park, MD 20742-4111, U.S.A.}
\affiliation{Joint Quantum Institute
University of Maryland, College Park, MD 20742-4111, U.S.A.}

\begin{abstract}
We show how the appearance of d-wave pairing in fermionic condensates manifests itself in inelastic light scattering. Specifically, we calculate the Bragg scattering intensity from the dynamic structure factor and the spin susceptibility, which can be inferred from spin flip Raman transitions. This information provides a precise tool with which we can identify nontrivial correlations in the state of the system beyond the information contained in the density profile imaging alone. Due to the lack of Coulomb effects in neutral superfluids, this is also an opportunity to observe the Anderson-Bogoliubov collective mode.
\end{abstract}



\date{\today}

\maketitle

\section{Introduction}

Ultra-cold atomic systems provide a way to build quantum simulators -- simulating a model with an experimental system~\cite{IBloch}. Studying models for strongly correlated systems can shed light on outstanding problems in condensed matter physics, for example whether the ground state of the doped Hubbard or t-J models supports d-wave superconductivity~\cite{Hofstetter}. It is important to have precise tools to identify and characterize the resulting phases. Establishing unconventional pairing symmetries requires new techniques in the setting of ultra-cold gases. Traditional probes like phase sensitive measurements, transport, or scanning tunneling microscopy, are all unavailable to condensates in optical systems. However, techniques do exist which provide information beyond the scope of what time-of-flight imaging can do, allowing precise identification of new phases.

We calculate the zero temperature response of a fermionic d-wave superfluid to inelastic light scattering. The primary emphasis in our work is that the inelastic response functions we provide are specific enough to unambiguously identify the underlying d-wave symmetry. We also have the opportunity to observe collective modes in a neutral paired superfluid. To this end, we also look at s-wave pairing. In solids, this mode is lifted to the plasma frequency by the Coulomb interaction and disorder can play an important role, making the collective mode difficult to observe. In the context of condensates, the complications from the Coulomb interaction and disorder are absent. A disadvantage is the effect of the confining potential which is usually present even for optical lattices.

Previous studies have focused on identifying s-wave superfluidity~\cite{ZhangSackettHulet,Combescot,Minguzzi,BuchlerZollerZwerger,OhashiGriffin}, or using noise-correlations to identify many-body states~\cite{AltmanDemlerLukin}, or using a periodic driving of the system then observing its response~\cite{Pekker}. Collective modes in chiral p-wave superconductors have also been examined~\cite{Yakovenko}. We explore the confluence of these ideas, focusing on identifying d-wave superfluids, and on the effect of their collective mode.

\section{Bragg Scattering}

We recall the textbook~\cite{BECbook} discussion of inelastic light probes here for clarity. Bragg scattering is a two photon process that transfers momentum $\Q=\k_1-\k_2$ and energy $\Omega=\omega_1-\omega_2$ to the ground state, and can be used to probe density fluctuations of the system. The probing light can be tailored to couple to density and between hyperfine degrees of freedom. The intensity of the light from the laser beams takes the form, $I_0 \cos(\Q \cdot\r- \Omega t)$, and the atoms in the sample will feel an optical potential due to the AC Stark effect. We label the couplings to the density (d) or spin (s) degree of freedom as $V_{d,s}$. The perturbation is taken as ~\cite{BECbook,Stamper-Kurn}
\begin{equation}
\hat{H}' =\sum_{d,s} \frac{V_{d,s}}{2}(\delta\hat{\rho}_{d,s,\Q}^{\dag}e^{-i\Omega t}+h.c.),
\label{pert}
\end{equation}
where  $\delta\hat{\rho}_{d,s,\Q}$ is the variation in the density or spin density. We will generalize this later in the paper.

The dynamic structure function $S(\Q,\Omega)$ is the Fourier transform of the density-density correlation function, where $\Q$ and $\Omega$ are respectively the momentum and energy transferred by the probe.  The excitation rate per particle is given by  $2\pi (\frac{V_0}{2\hbar})^2 S(\Q,\Omega)$ ~\cite{Stamper-Kurn}. The dynamic structure factor can be measured in more than one way~\cite{Stamper-Kurn,HuletPaper}. For example, after the atomic sample is illuminated by the two laser beams, the trap is switched off, and time of flight images can measure the Bragg scattered atoms -- distinguished because they are displaced by the absorbed momentum. What is actually measured in this experiment is the rate of momentum transfer ~\cite{MomentumBragg}, which is $\frac{d \bf{P}}{dt}= \Q (\frac{V_0}{2})^2 \frac{2 \pi}{\hbar} [S(\Q,\Omega)-S(-\Q,-\Omega) ]\propto (1-e^{-\beta \Omega})S(\Q,\Omega)\rightarrow \lim_{T=0}S(\Q,\Omega)$. In the end, Bragg scattering provides a measurement of the imaginary part of the density-density correlation function which we calculate. Similarly, it has been suggested that the spin flip rate observed by Raman transitions can be used to study the spin susceptibility~\cite{BuchlerZollerZwerger,HuletPaper}. 
These two probes are specific enough to confirm unconventional Fermi superfluid pairing when it arises in condensates.

\section{Hamiltonian and Response Functions}

\begin{figure}[b]
\includegraphics[width=0.3\textwidth]{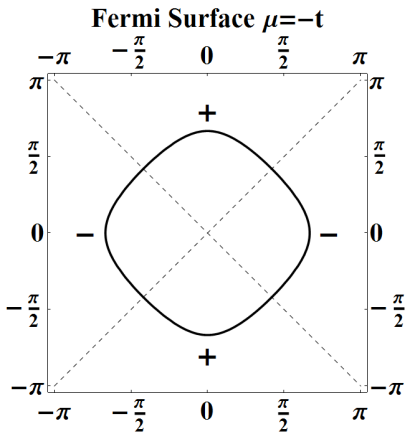}
\caption[Fermi Surface]{The Fermi surface for a square lattice in the tight binding limit including only nearest neighbor hoppings at $\mu=-t$. The nodal lines and sign of the gap are marked. }
\label{fermisurface}
\end{figure}

We limit our discussion to a dilute Fermi gas which undergoes pairing in the BCS limit. The BCS Hamiltonian generalized for anisotropic pairing is:
\begin{equation}
H=\sum_k \begin{array}{cc} (\hat{c}^{\dag}_{\k\uparrow} & \hat{c}_{-\k\downarrow}) \end{array}
\left( \begin{array}{cc}
  \xi_{\k} & \Delta_{\k}  \\
\Delta_{\k} &   - \xi_{\k}  \\
\end{array} \right) \left(
\begin{array}{c} \hat{c}_{\k\uparrow} \\
 \hat{c}^{\dag}_{-\k\downarrow} \end{array}\right).
\end{equation}

\begin{figure}[t]
\centering
\begin{tabular}{cc}
\multicolumn{1}{l}{\mbox{(a)}} &
	\multicolumn{1}{l}{\mbox{(b)}} \\
\includegraphics[width=.5\columnwidth,angle=0]{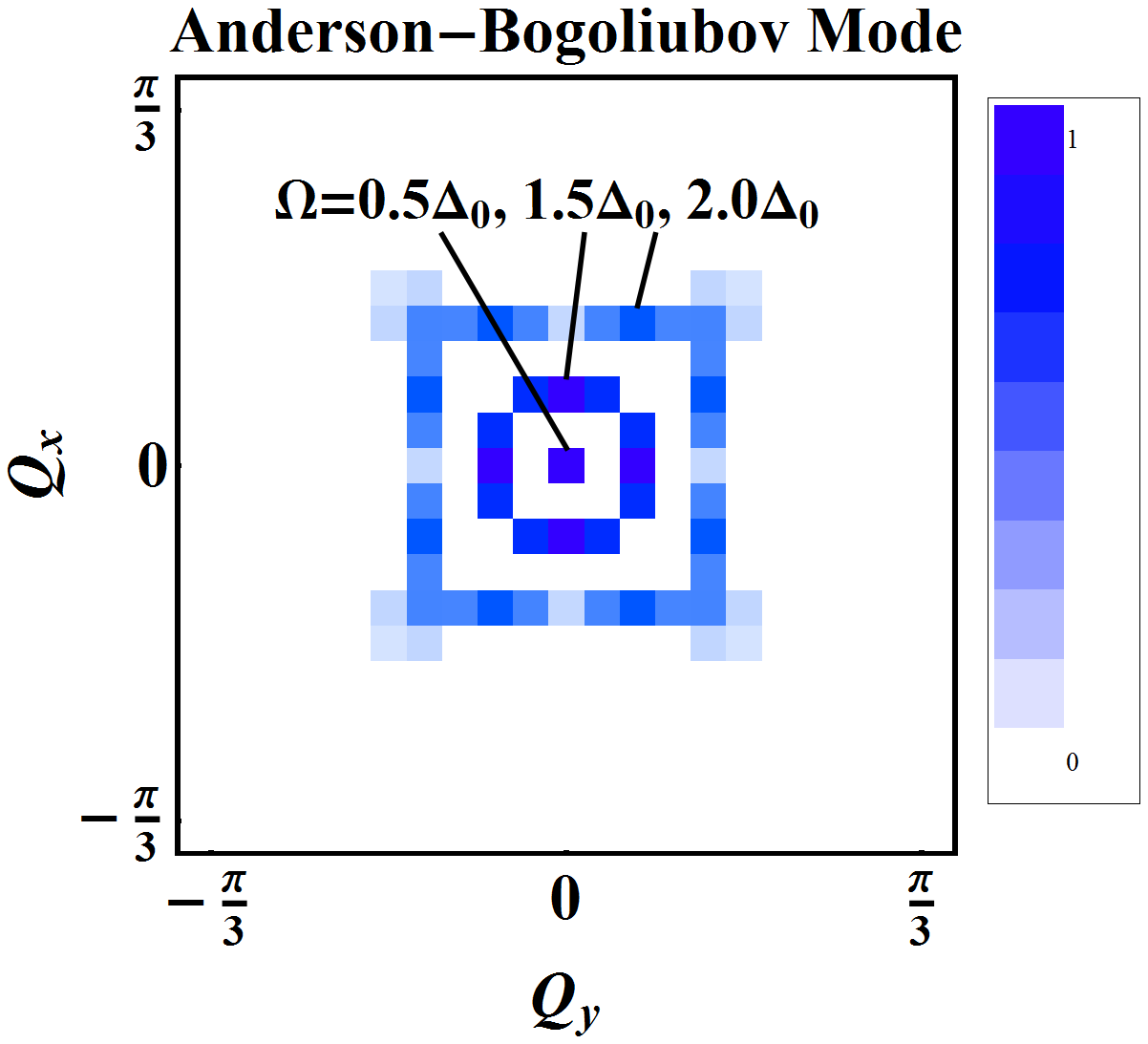} &
\includegraphics[width=.415\columnwidth,angle=0]{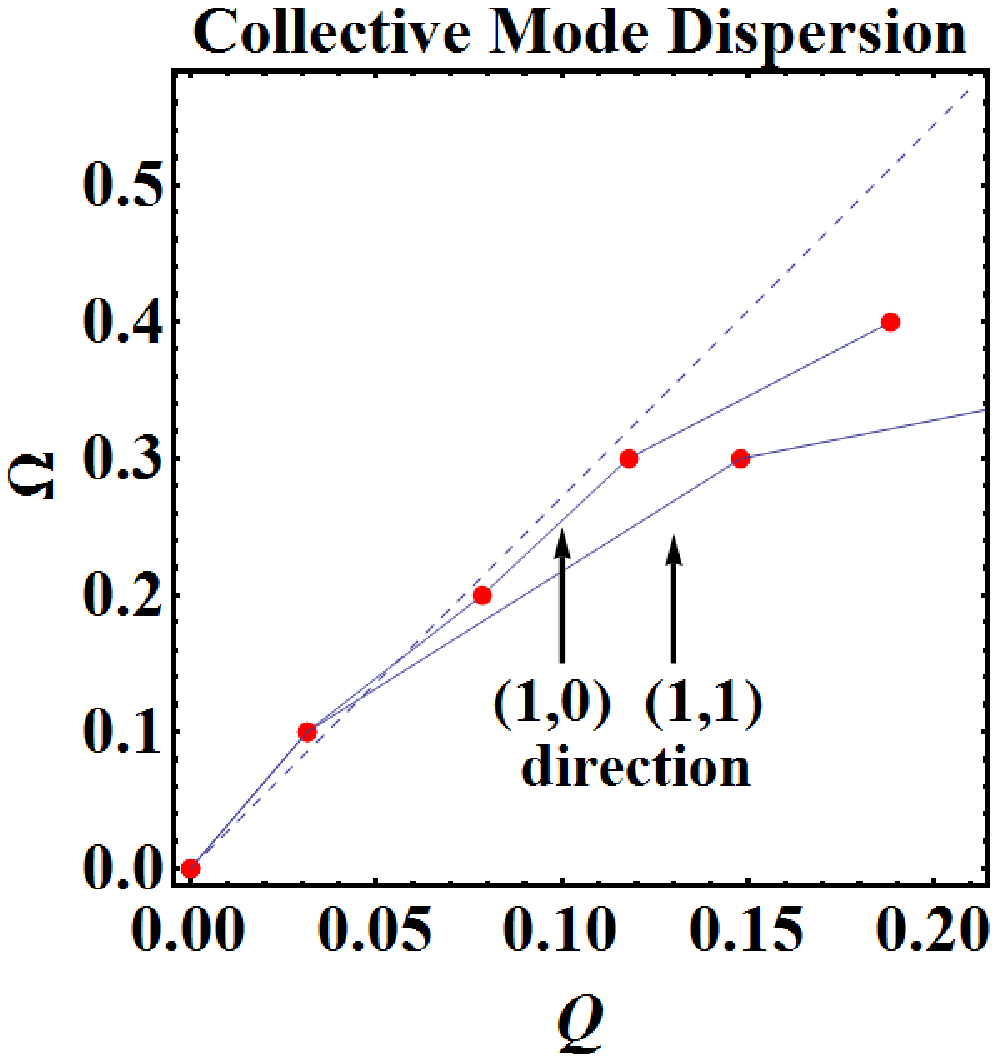}
\end{tabular}
\caption[Susceptibilities]{ (Color Online) Plot of $\rm{Im}[\Pi_{22}(0,0)-\Pi_{22}(\Q,\Omega)]^{-1}$ in the s-wave state  for (a) $\Omega=0.5\Delta_0$ (b) $\Omega=1.5\Delta_0$ (c) $\Omega=2.0\Delta_0$. The poles in the denominator of the second term in Eq.~(\ref{RPAresp}) illustrate the Anderson-Bogoliubov mode. Panel (b) shows the approximate dispersion captured from panel (a) in the (1,0) and (1,1) directions compared to the result $\Omega=\frac{v_F }{\sqrt{2}}Q$.}
\label{NormalSwave}
\end{figure}

For a two dimensional square lattice in the tight binding approximation, the dispersion is
\begin{equation}
\xi_{\k}=-2t[\cos(k_x)+\cos(k_y)]-\mu.
\end{equation}
The d-wave superconducting gap is
\begin{equation}
\Delta_{\k}=\Delta_0 \chi_{\k}, \qquad \chi_{\k}=\cos(k_x)-\cos(k_y).
\end{equation}

We chose the chemical potential to be $\mu=-t$, and $\Delta_0=0.2t$. The resulting Fermi surface is shown in Fig.~\ref{fermisurface}. We note that particle-hole symmetry takes $\mu=-t \rightarrow \mu=t$ and is just a translation of the Fermi surface by $(\pi,\pi)$. The Bogoliubov quasiparticles have an excitation spectrum given by: $E_{\k}=\sqrt{\xi_{\k}^2+\Delta_{\k}^2}$.
The Green's function for a d-wave superconductor takes the well known form in Nambu space,
\[
 \check{G}(\k,i \omega_n) =\left(i \omega_n \check{1}+ \xi_{\k} \check{\tau}_3+ \Delta_{\k} \check{\tau}_1 \right)^{-1} =\]
\begin{equation}
 \frac{1}{(i \omega_n)^2 - (\xi_{\k}^2 + \Delta_{\k}^2)}  \left(
\begin{array}{cc}
i \omega_n + \xi_{\k} & \Delta_{\k}  \\
\Delta_{\k} & i \omega_n - \xi_{\k}  \\
\end{array} \right).
\label{green}
\end{equation}

In this work, the inclusion of both the d-wave quasiparticle and collective response is crucial. Following the pioneering works of Kulik et al.~\cite{KEO}, Wong and Takada~\cite{WongTakada}, and Ohashi and Takada~\cite{OhashiTakada,OhashiTakada2}, we calculate the physical response by including fictitious interactions. To accomplish this we generalize the density operators.  The Pauli matrices in Nambu space will be denoted by $\check{\tau}_i$ with $\check{\tau}_0=1$. $\hat{\Psi}^{\dag}_{\k}=\begin{array}{cc} (\hat{c}^{\dag}_{\k\uparrow} & \hat{c}_{-\k\downarrow}) \end{array}$ is a spinor in Nambu space. In this notation, the generalized densities are
\[\check{\rho}_{i\Q}=  \sum_{\k} \gamma_{\k+\Q/2,i}\hat{\Psi}^{\dag}_{\k}\check{\tau}_i \hat{\Psi}_{\k+\Q},\;\, \gamma_{\k,i=0,1,2,3}=(1,\chi_{\k},\chi_{\k},1),
\]
\begin{eqnarray}
\check{\rho}_{0\Q}=  \sum_{\k}  \hat{c}^{\dag}_{\k,\uparrow} \hat{c}_{\k+\Q,\uparrow} - \hat{c}^{\dag}_{\k,\downarrow} \hat{c}_{\k+\Q,\downarrow}\\
\check{\rho}_{1\Q}=  \sum_{\k} \chi_{\k+\Q/2}\left(\hat{c}^{\dag}_{\k\uparrow}\hat{c}^{\dag}_{-\k-\Q\downarrow}+h.c.\right)\\
\check{\rho}_{2\Q}=  \sum_{\k} \chi_{\k+\Q/2}\left(i\hat{c}^{\dag}_{\k\uparrow}\hat{c}^{\dag}_{-\k-\Q\downarrow}+h.c.\right) \\
\check{\rho}_{3\Q}=  \sum_{\k} \hat{c}^{\dag}_{\k,\uparrow} \hat{c}_{\k+\Q,\uparrow}+ \hat{c}^{\dag}_{\k,\downarrow} \hat{c}_{\k+\Q,\downarrow}.
\end{eqnarray}

Physically, $\check{\rho}_{3\Q}=\check{\rho}_{d\Q}$ is the density operator, while $\check{\rho}_{1 \Q}$ and $\check{\rho}_{2 \Q}$ can be thought of as, respectively, qualitatively encapsulating the amplitude and phase of the order parameter. The spin-density, treated separately throughout the paper, is $\check{\rho}_{s\Q}=\check{\rho}_{0\Q}$. By treating the spin density separately we do not include the possibility of spin-phase, spin-amplitude, etc.~modes. According to Buchler et al. \cite{BuchlerZollerZwerger}, the contribution from collective terms for $Im[\Pi_{00}]$ vanish, so $Im[\Pi_{00}]=Im[\Pi^0_{00}]$. In Zou et al.\cite{BraggZouHuiHu} a self-consistent spin-spin interaction is included in the Random Phase Approximation for the s-wave case which goes beyond our analysis. For interactions, we include the terms:
\begin{equation}
\hat{H}_{int}=\frac{-g}{2}(\check{\rho}_{1\Q}\check{\rho}_{1,-\Q}+\check{\rho}_{2\Q}\check{\rho}_{2,-\Q}).
\end{equation}

The density and spin response of the system is calculated using linear response theory. The density-density response has been generalized to be a matrix of density responses in Nambu space. The generalized response functions,
\begin{equation}
\Pi^0_{ij}=-\int_0^{\beta} d \tau e^{i \Omega_m \tau} \langle T_{\tau}\{\check{\rho}_{i\Q}(\tau)\check{\rho}_{i-\Q}(0)\}\rangle,
\end{equation} take the form:
\begin{align}
\nonumber
& \Pi^0_{ij}(\Q,\Omega_m)=\int \frac{d^d \k}{(2 \pi)^d} \gamma_{\k+\Q/2 i} \gamma_{\k+\Q/2 j}\sum_{\omega_n}\\
&\times \frac{\rm{Tr}}{2 \beta} ~[\check{\tau}_i \check{G}(\k,\omega_n)\check{\tau}_j \check{G}(\k+\Q,\omega_n + \Omega_m)].
\label{Pi0}
\end{align}

The gap equation at zero temperature,
\begin{equation}
\Delta_{\k} = - \sum_{\k'} V(\k,\k') \frac{\Delta_{\k'}}{2 E_{\k'}},\end{equation}

is equivalent to $0=1+\frac{g}{2}\Pi^0_{22}(0,0)$ where ``g" is the BCS coupling constant stemming from assuming a separable pairing interaction $V(\k,\k')=g \chi_{\k} \chi_{\k'}$. Zero temperature Bragg scattering would correspond to a measurement of $\rm{Im} \, \Pi_{33}(\Q,\Omega)$, and a different experiment could measure $\rm{Im} \,\Pi_{00}(\Q,\Omega)$, the spin susceptibility.

\begin{figure*}[t]
\begin{tabular}{ccc}
\multicolumn{1}{l}{\mbox{(a)}} &
	\multicolumn{1}{l}{\mbox{(b)}} & 	\multicolumn{1}{l}{\mbox{(c)}} \\
\includegraphics[width=0.25\textwidth,angle=0]{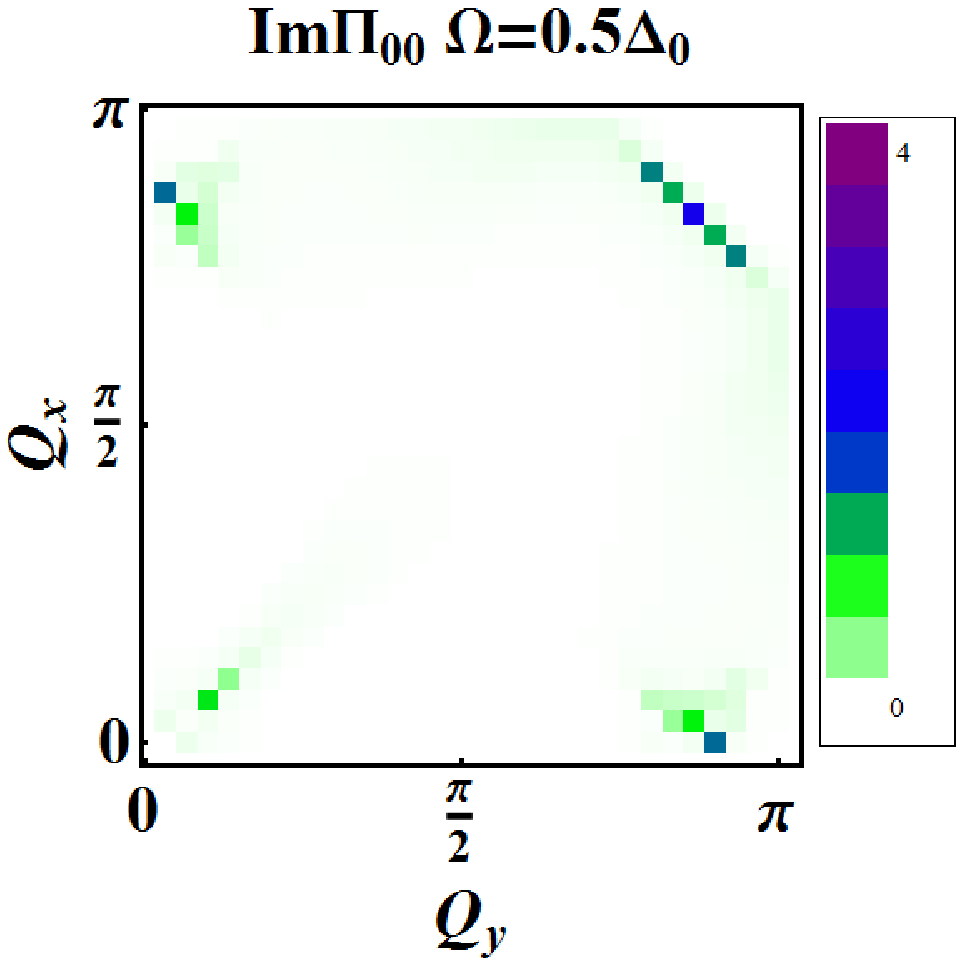} &
\includegraphics[width=0.25\textwidth,angle=0]{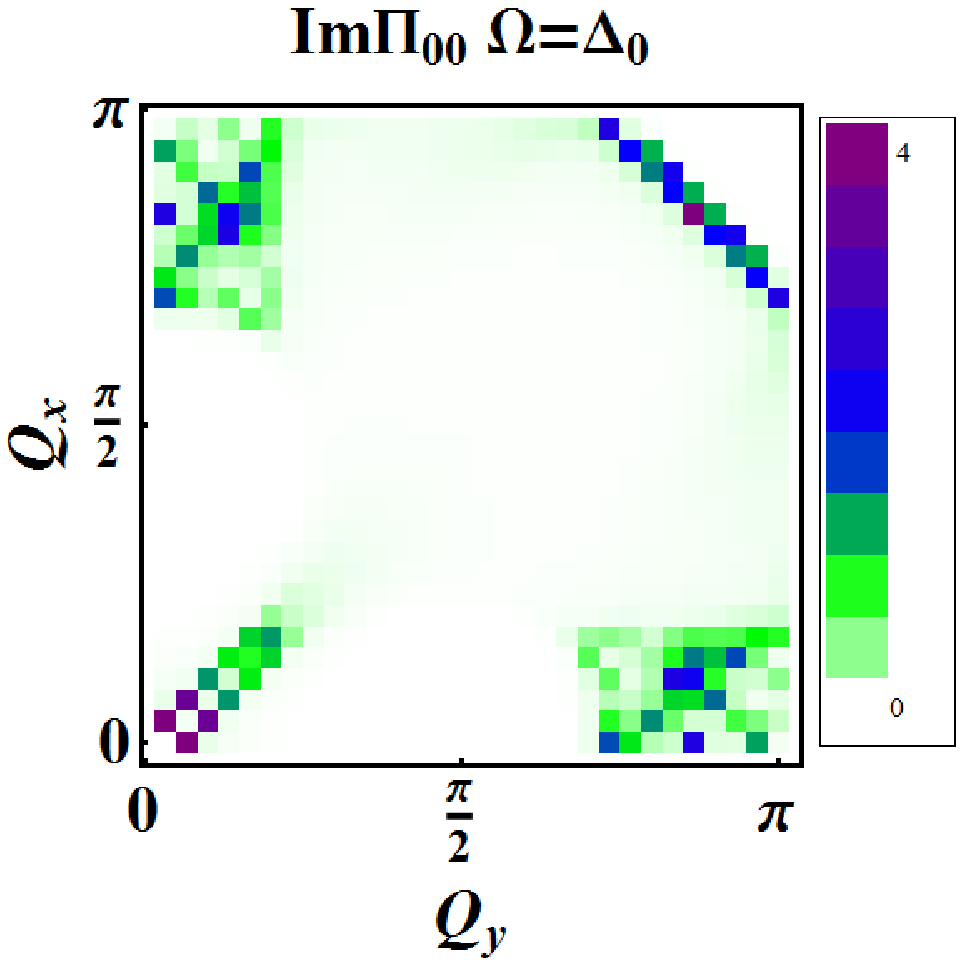} &
\includegraphics[width=0.25\textwidth,angle=0]{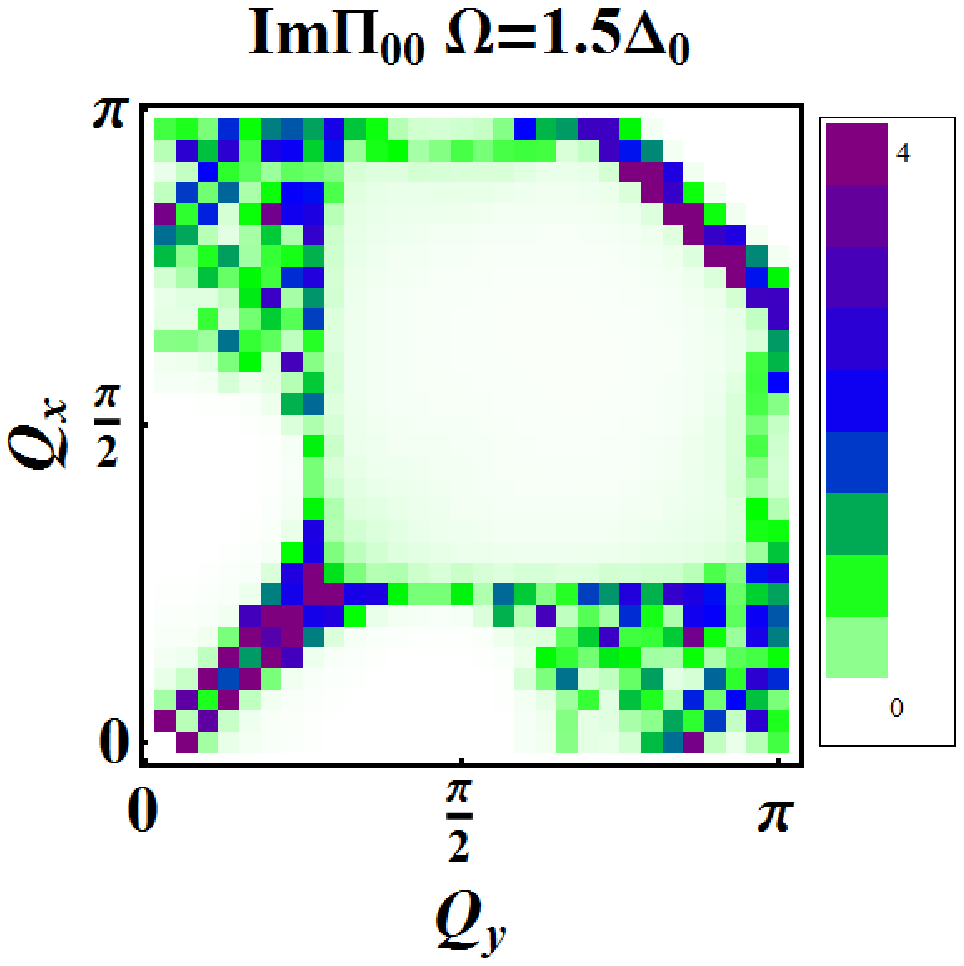} \\ 
\end{tabular}
\caption[]{(Color Online) The imaginary part of the zero temperature spin susceptibility, $\rm{Im} [\Pi^0_{00}]$, Eq.~\ref{Pi0} in the d-wave state, for frequencies $(a)\Omega=0.5\Delta_0,\; (b)\Delta_0,\; (c) 1.5\Delta_0$, with $\Delta_0=0.2t$. Notice that there is zero response at $\Q=(0,0)$ consistent with the fact that the spin structure factor vanishes in the zero momentum limit \cite{BuchlerZollerZwerger}.}
\label{SpinSusc}
\end{figure*}

\begin{figure*}[t!]$
\begin{array}{ccc}
\multicolumn{1}{l}{\mbox{(a)}} &
	\multicolumn{1}{l}{\mbox{(b)}} & 	\multicolumn{1}{l}{\mbox{(c)}} \\
\includegraphics[width=0.25\textwidth,angle=0]{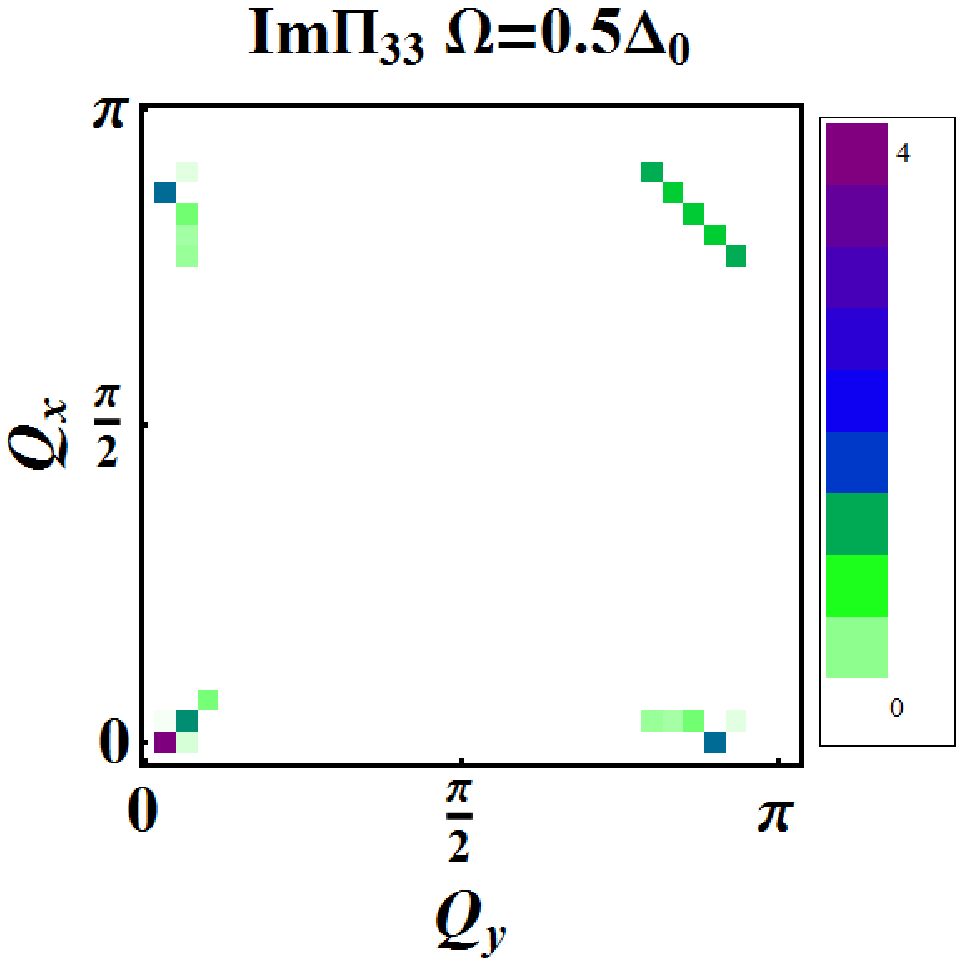} &
\includegraphics[width=0.25\textwidth,angle=0]{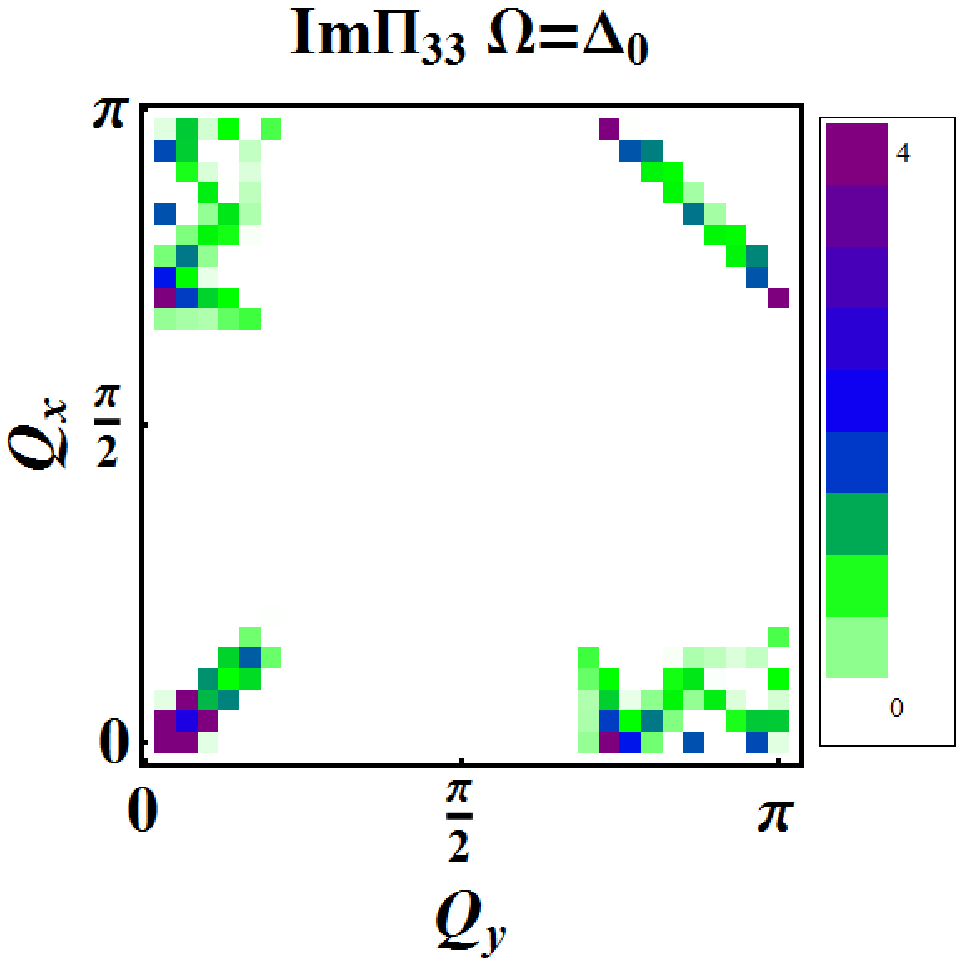} &
\includegraphics[width=0.25\textwidth,angle=0]{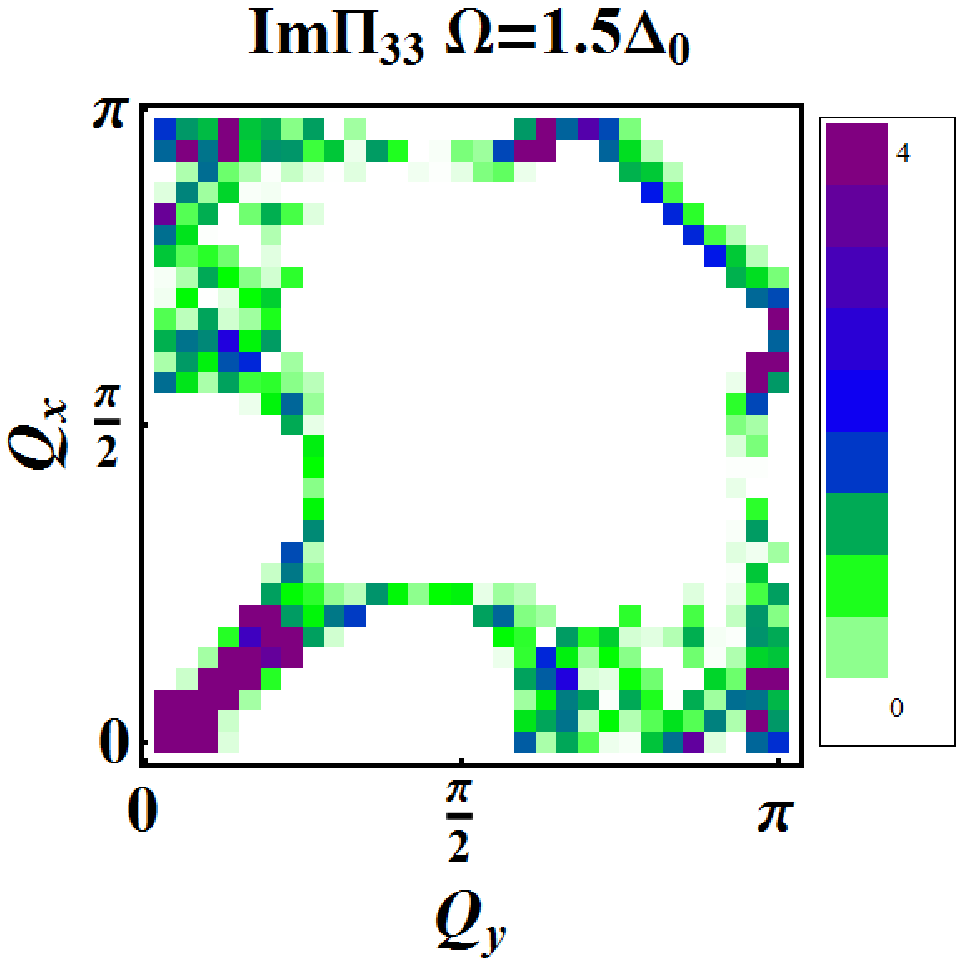}
\end{array}$
\caption[]{(Color Online) The imaginary part of the zero temperature density-density response, $\rm{Im}[\Pi_{33}]$, Eq.~\ref{RPAresp} in the d-wave states, for frequencies  $(a) \Omega=0.5\Delta_0,\; (b) \Delta_0,\; (c) 1.5\Delta_0$. We have artificially put a mark in the center of Figs (a) and (c) where the data was clipped to represent points off the scale. For these plots $\Delta_0=0.2t$. }
\label{densityResp}
\end{figure*}

The density-density response at T=0 comes from the following integrals
\begin{align}
\nonumber& \Pi^{ab}_{ij}(\Q,\Omega) = i \int \frac{d^2 \k}{(2 \pi)^2} d\omega
\nonumber  \gamma_{\k+\Q/2, i} \gamma_{\k+\Q/2, j}  \\
& \times \frac{\omega (\omega + \Omega)+a\xi_{\k} \xi_{\k+\Q} + b \Delta_{\k} \Delta_{\k+\Q}}{(\omega^2-E_{\k}^2)[(\omega+\Omega)^2-E_{\k+\Q}^2 ]}.
\end{align}
The upper indices are $a,b=\pm 1$ and are determined by which Nambu-space matrices occur in the lower indices for the generalized densities. The spin susceptibility ($i=0$ $j=0$) corresponds to $\rm{a}=1$, $\rm{b}=1$ and has been previously calculated~\cite{MazinYakovenko}. The density channel ($i=3$ $j=3$) has $\rm{a}=1$, $\rm{b}=-1$ and the phase channel ($i=2$ $j=2$) has $\rm{a}=-1$, $\rm{b}=-1$.  After completing the contour integral, it is helpful to replace the terms $E(E+\Omega)$ by a choice which cancels some terms in the denominators, $E(E+\Omega)=E(E+E'+\Omega)-E E'$.

\begin{align}
\nonumber &\Pi^{ab}_{ij}(\Q,\Omega)=
-\int \frac{d^2 \k}{(2 \pi)^2}
\frac{\gamma_{\k+\Q/2, i} \gamma_{\k+\Q/2, j}}{4} \\ \label{Pi0ijab}
&\times \left(1-a\frac{\xi_\k \xi_{\k+\Q}}{E_\k E_{\k+\Q}}-b\frac{\Delta_\k \Delta_{\k+\Q}}{E_\k E_{\k+\Q}}\right) \\
\nonumber
&\times\left(\frac{1}{\Omega+E_{\k+\Q}+E_{\k}+i\eta}+\frac{1}{E_{\k+\Q}+E_{\k}-\Omega+i\eta}\right).
\end{align}

The phase-density response function $\Pi_{32}=-\Pi_{23}$ is also needed:
\begin{eqnarray}
\Pi_{32}(\Q,\Omega)=2i \int \frac{d^2 \k}{(2 \pi)^2}  \gamma_{\k+\frac{\Q}{2}, i} \frac{\Delta_\k E_{\k+\Q} + E_{\k }\Delta_{\k+\Q}}{4{E_{\k} E_{\k+\Q}}} \\
\nonumber
\times
 \left( \frac{1}{\Omega+E_{\k+\Q}+E_{\k}+i\eta}-\frac{1}{E_{\k+\Q}+E_{\k}-\Omega+i\eta} \right).
\end{eqnarray}

\begin{figure}[t]
\begin{tabular}{cc}
\multicolumn{1}{l}{\mbox{(a)}} &
	\multicolumn{1}{l}{\mbox{(b)}} \\
\includegraphics[width=0.24\textwidth,angle=0]{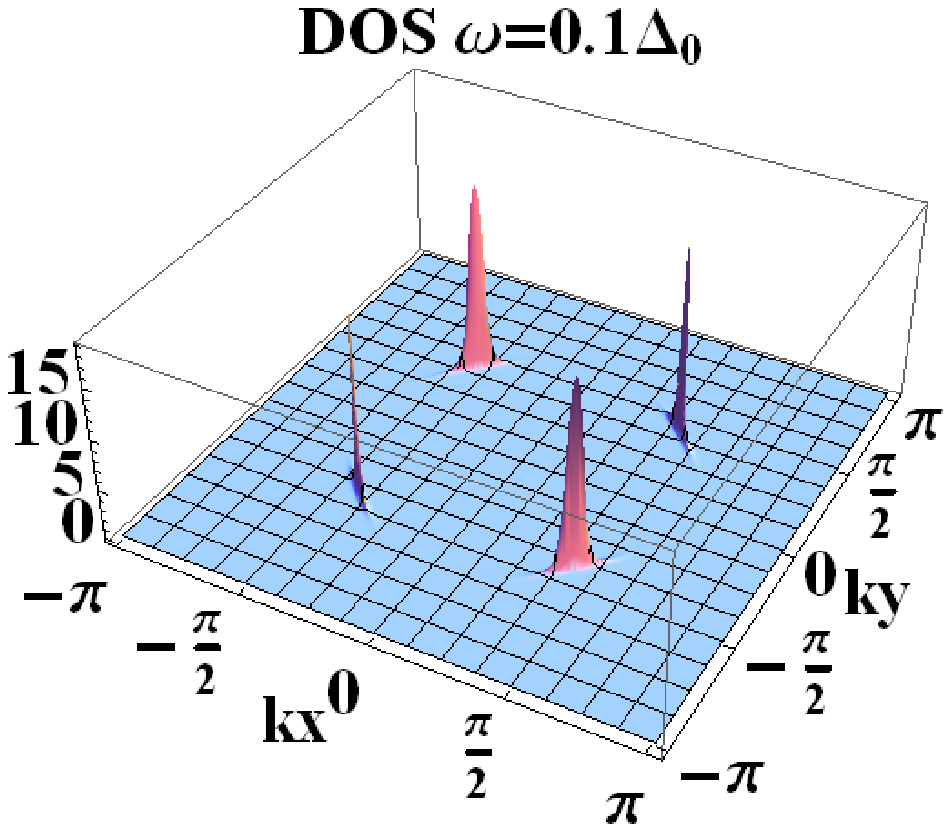} &
\includegraphics[width=0.24\textwidth,angle=0]{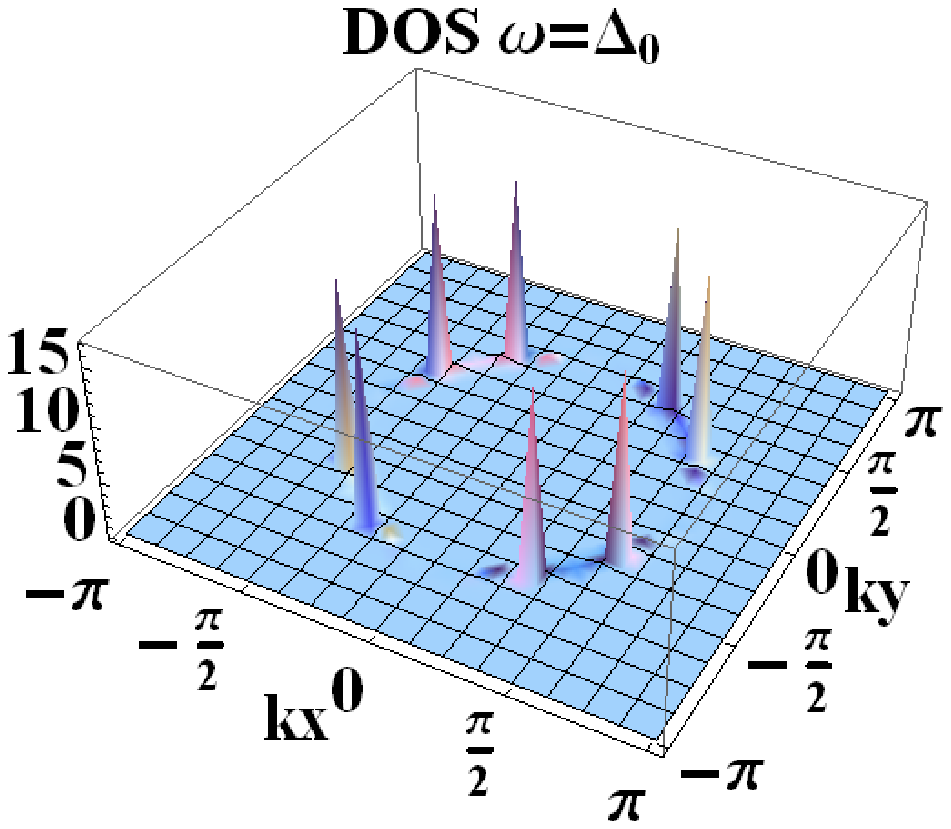} \\
\multicolumn{1}{l}{\mbox{(c)}} &
	\multicolumn{1}{l}{\mbox{(d)}} \\
\includegraphics[width=0.24\textwidth,angle=0]{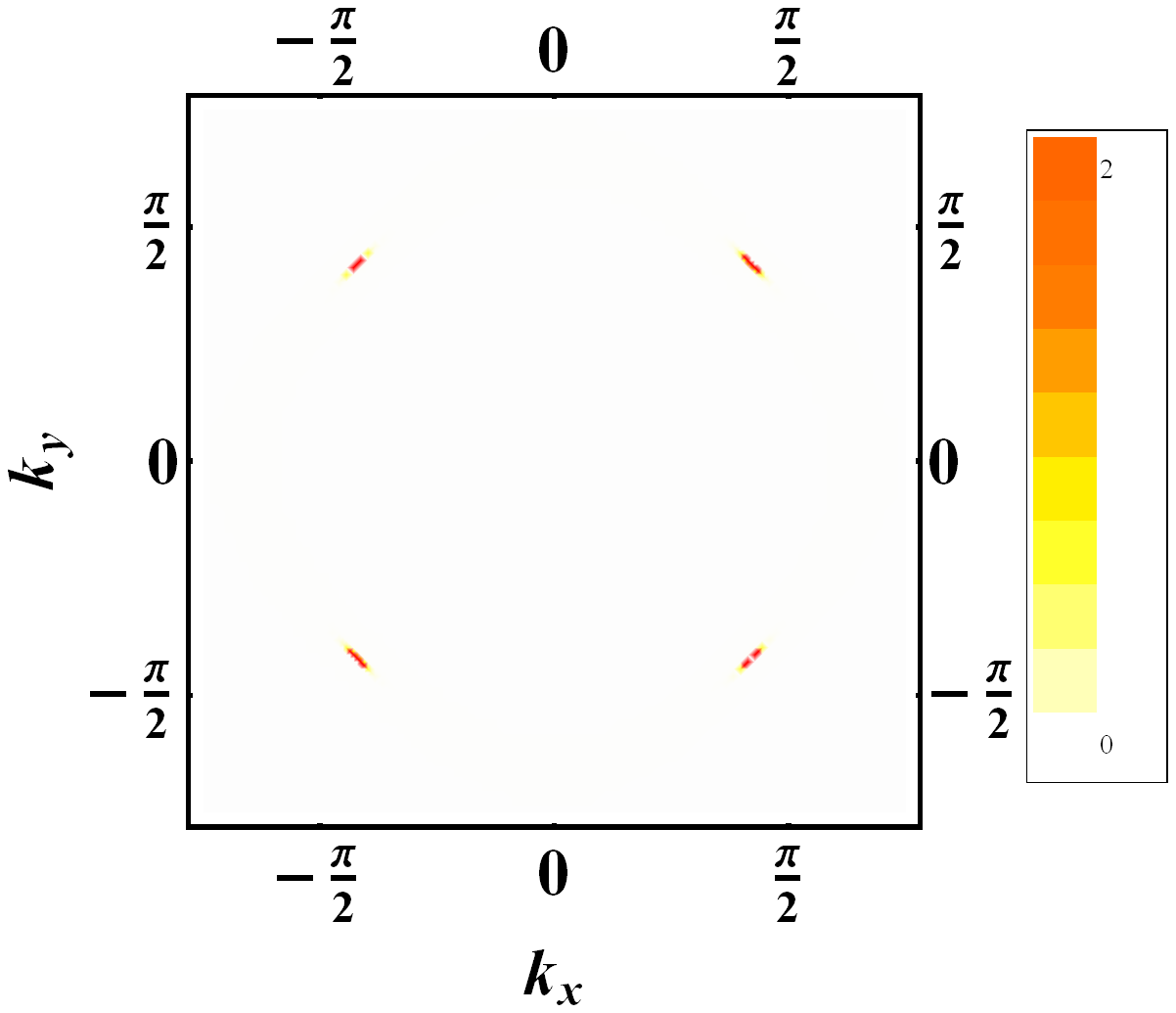} &
\includegraphics[width=0.24\textwidth,angle=0]{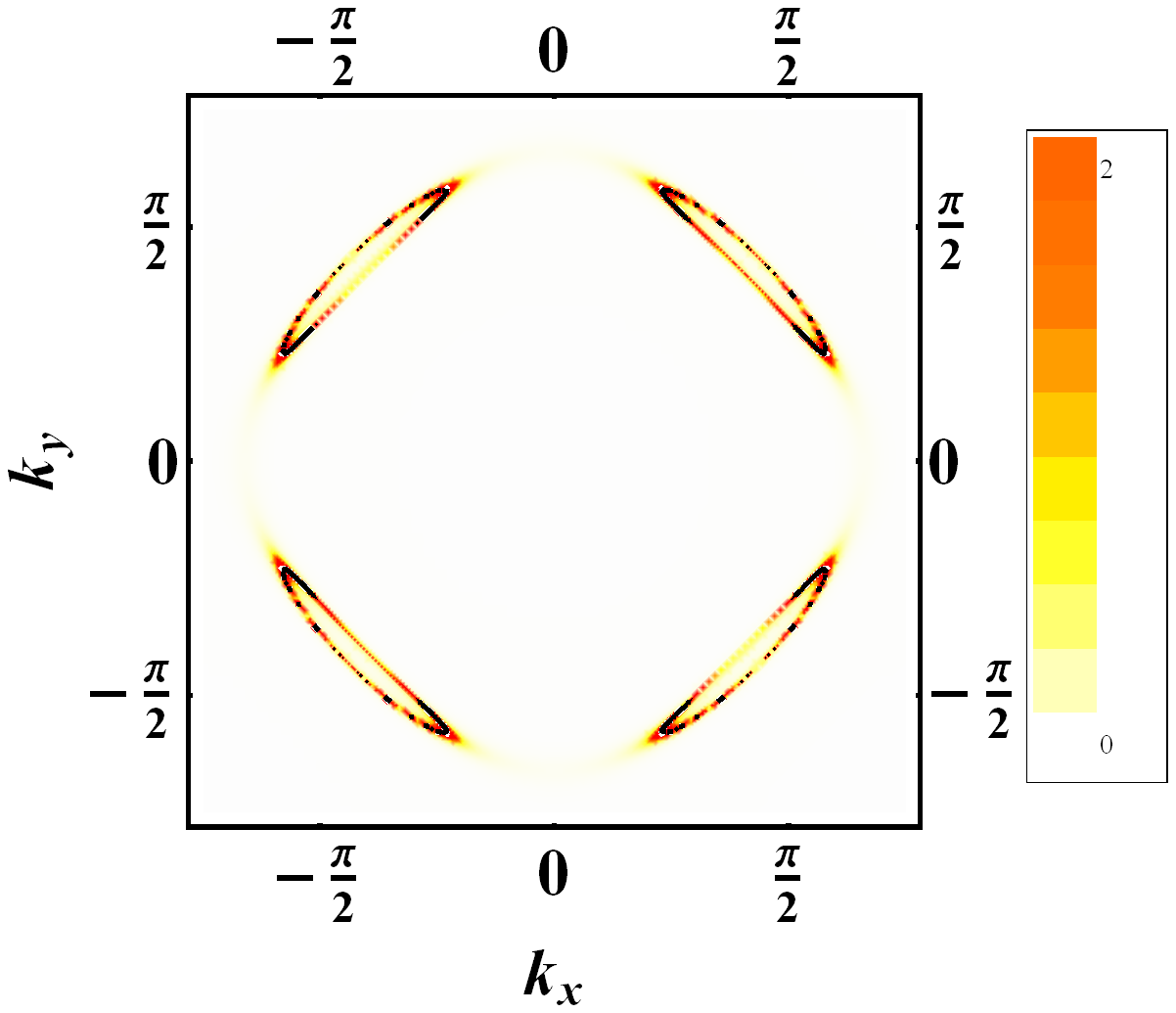}
\end{tabular}
\caption[]{(Color Online) Momentum resolved density of states for fixed frequency $\Omega$. For $\Omega=0.1\Delta_0$(a) we see nodal excitations. For $\Omega=\Delta_0$ (b) the region of highest DOS corresponds minima in $\nabla_{\k} E_{\k}$. The solid black line is the equal energy contour for the same energies.}
\label{dwaveBZ}
\end{figure}

The Feynman diagrams of the random phase approximation (RPA) are resummed for a neutral superfluid. It is found by explicit calculation\cite{KEO,WongTakada,OhashiTakada} that the amplitude modes decouple from the density and phase oscillations, $\Pi_{12}=\Pi_{13}=0$, so the the density response is:
\begin{equation}
\left(\begin{array}{cc}
\Pi_{22} & \Pi_{23}  \\
\Pi_{32} &  \Pi_{33}  \\
\end{array} \right)^{-1}
=\left(\begin{array}{cc}
\Pi^0_{22} & \Pi^0_{23}  \\
\Pi^0_{32} &  \Pi^0_{33}  \\
\end{array} \right)^{-1}-\left(\begin{array}{cc}
-g/2 & 0  \\
0 &  0  \\
\end{array} \right). \end{equation}

The full density-density response~\cite{WongTakada}, after analytic continuation $\Omega_m \rightarrow \Omega + i\eta$ (for numerics $\eta$ was $10^{-3}t$),
\begin{equation}
\Pi_{33}(\Q,\Omega)=\Pi^0_{33}(\Q,\Omega)-\frac{\Pi^0_{32}(\Q,\Omega)\Pi^0_{23}(\Q,\Omega)}{\frac{2}{g}+\Pi^0_{22}(\Q,\Omega)}.
\label{RPAresp}
\end{equation}

Note that the gap equation is used to eliminate g in terms of $\Pi^0_{22}(0,0)$.

The formula in Eq.(\ref{RPAresp}) has a clear interpretation. The first term is the ordinary density-density response. The second term is a coupling between the density and phase response with a phase mode in the denominator. A pole in the second term is the Goldstone, or Anderson-Bogoliubov, mode for the neutral superfluid.

The finite temperature expression is required to obtain the normal state limit. This T=0 result only contains the pair-breaking terms, because the scattering terms vanish for T=0, and both are necessary to recover the normal state limit, which follows directly from the result~\cite{PhysRevLett.68.125}.

The calculation of the inelastic neutron response, notably in cuprates~\cite{KLevinNeutron,PhysRevB.49.4235}, is very similar to this calculation; frequently, in cuprate calculations~\cite{PhysRevLett.75.4130,PhysRevB.47.9124} the RPA analysis contains a magnetic structure $J(\q)=[\cos (q_x) + \cos (q_y)]$, bilayer effects, or Coulomb interactions which are absent here.

\section{Discussion}

\begin{figure}[t]
\begin{tabular}{cc}
\multicolumn{1}{l}{\mbox{(a)}} &
	\multicolumn{1}{l}{\mbox{(b)}} \\
\includegraphics[width=0.22\textwidth,angle=0]{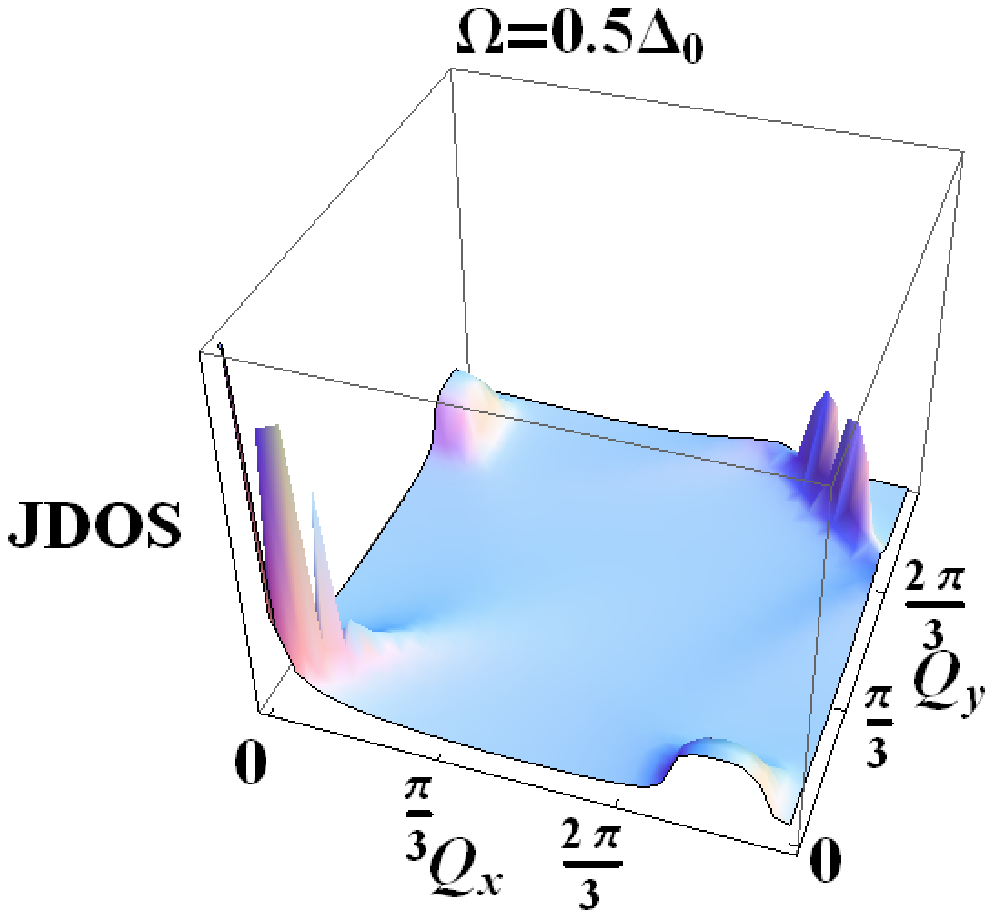}  &
\includegraphics[width=0.22\textwidth,angle=0]{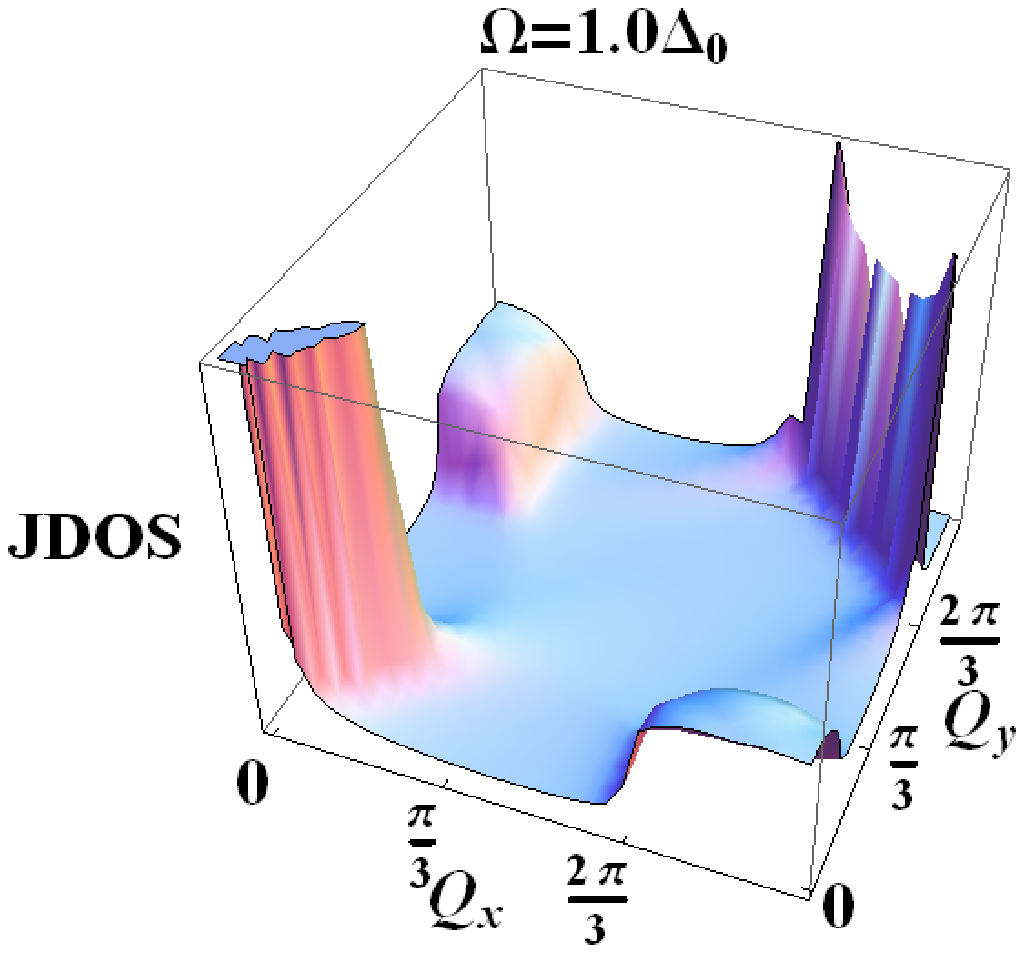} \\
\multicolumn{1}{l}{\mbox{(c)}} &
	\multicolumn{1}{l}{\mbox{(d)}} \\
\includegraphics[width=0.23\textwidth,angle=0]{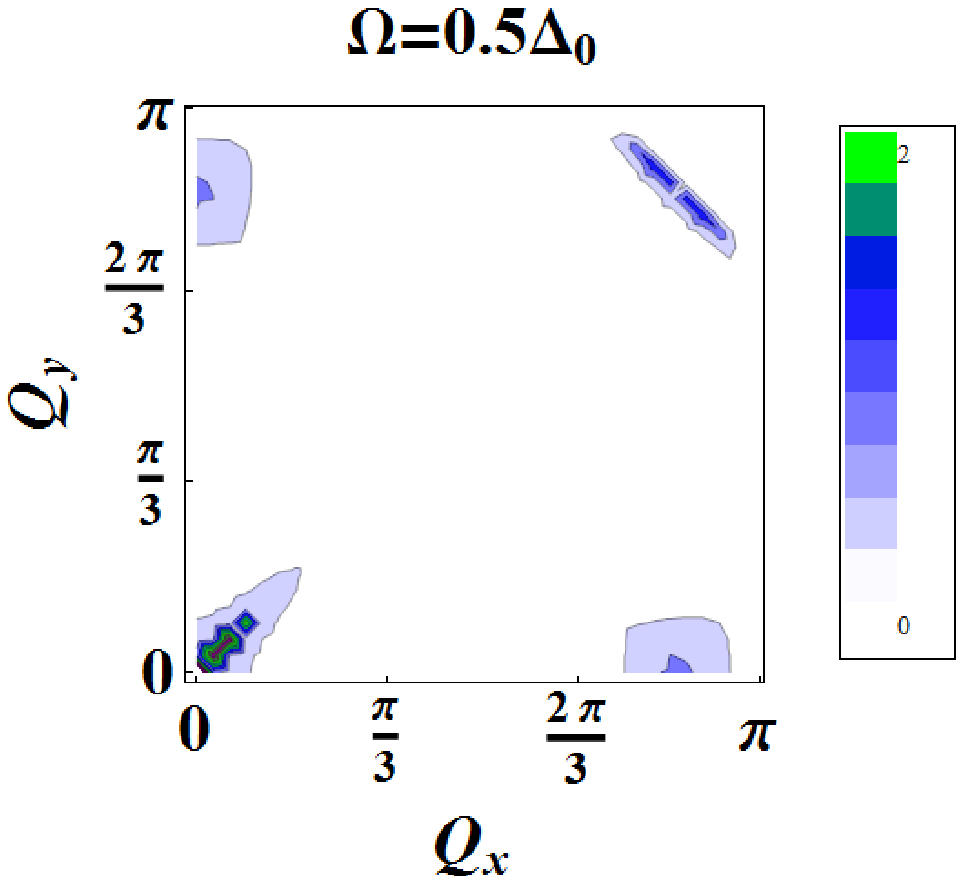} &
\includegraphics[width=0.23\textwidth,angle=0]{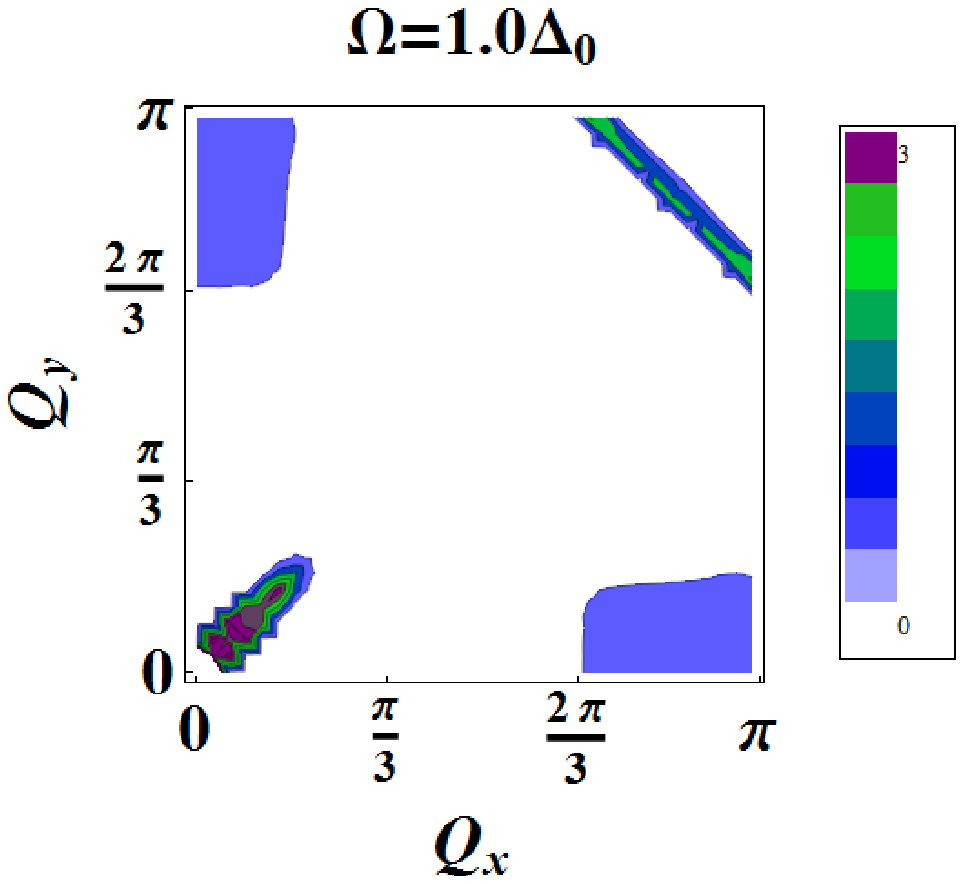}
\end{tabular}
\caption[]{(Color Online) Joint density of states, $\rm{JDOS}(\Q,\Omega)$, for two-particle excitations over the range (0,$\pi$) (a,b) and contour plots of the same (c,d).}
\label{JDOS}
\end{figure}

If the system were to condense into an s-wave superfluid, we should see the effect of opening a gap on the Fermi surface. The only low-energy feature present in the s-wave case is the Anderson-Bogoliubov mode. All non-Goldstone excitations are gapped below the scale $\Omega<2\Delta_0$. In Fig.~\ref{NormalSwave}a we show the remaining gapless collective excitations. Maxima in the signal for various frequencies are compared to the exact result $\Omega=\frac{v_F }{\sqrt{2}}Q$ \cite{RanderiaABdispersion} in Fig.~\ref{NormalSwave}b. The presence of a square lattice dispersion causes anisotropy in the Anderson-Bogoliubov mode's dispersion near the pair breaking frequency $\Omega=2\Delta_0$ shown in Fig.~\ref{NormalSwave}. In ordinary s-wave superconductors, the presence of the Coulomb interaction is said to lift the energy of this excitation to the plasma scale. In a neutral superfluid, the absence of a Coulomb interaction and gapped quasiparticle excitations means that the only low energy mode is this collective mode.

\begin{figure*}
\begin{tabular}{ccc}
\multicolumn{1}{l}{\mbox{(a)}} &
	\multicolumn{1}{l}{\mbox{(b)}} & 	\multicolumn{1}{l}{\mbox{(c)}} \\
\includegraphics[width=0.31\textwidth,angle=0]{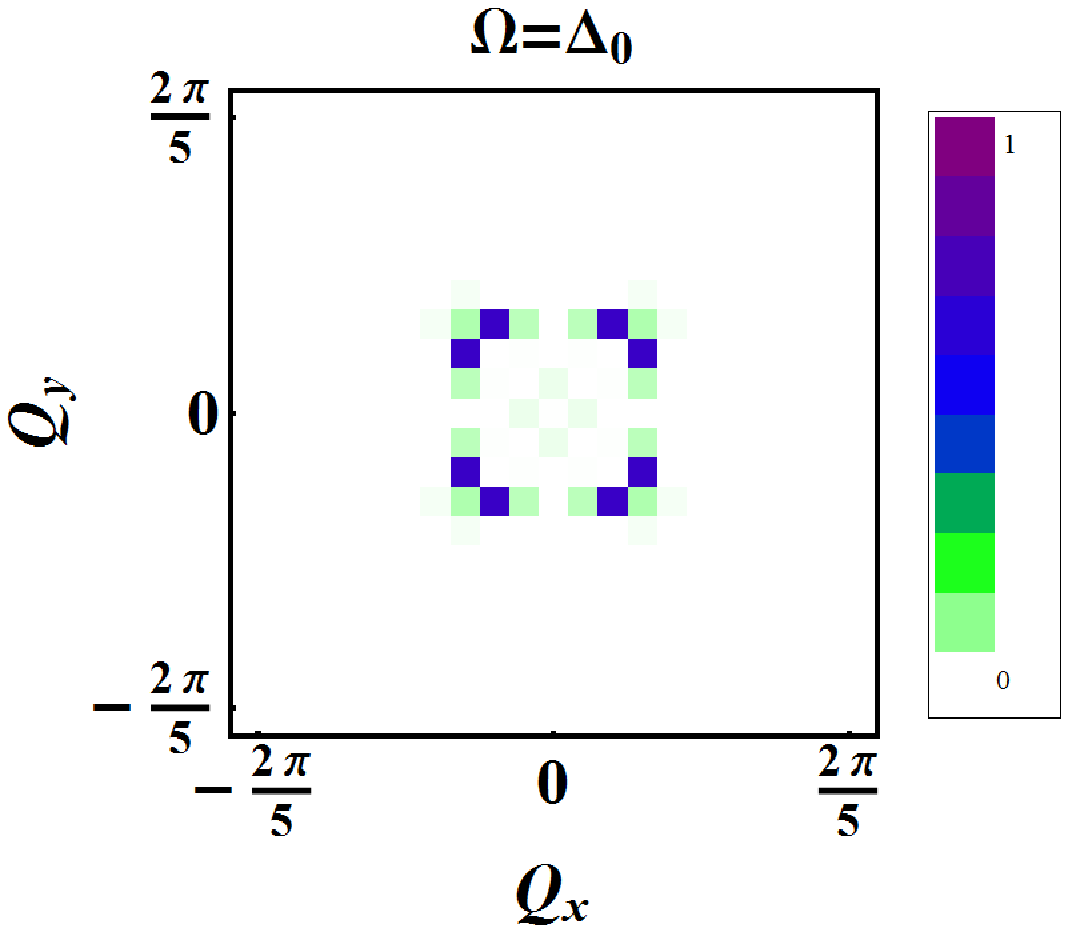}  &
\includegraphics[width=0.31\textwidth,angle=0]{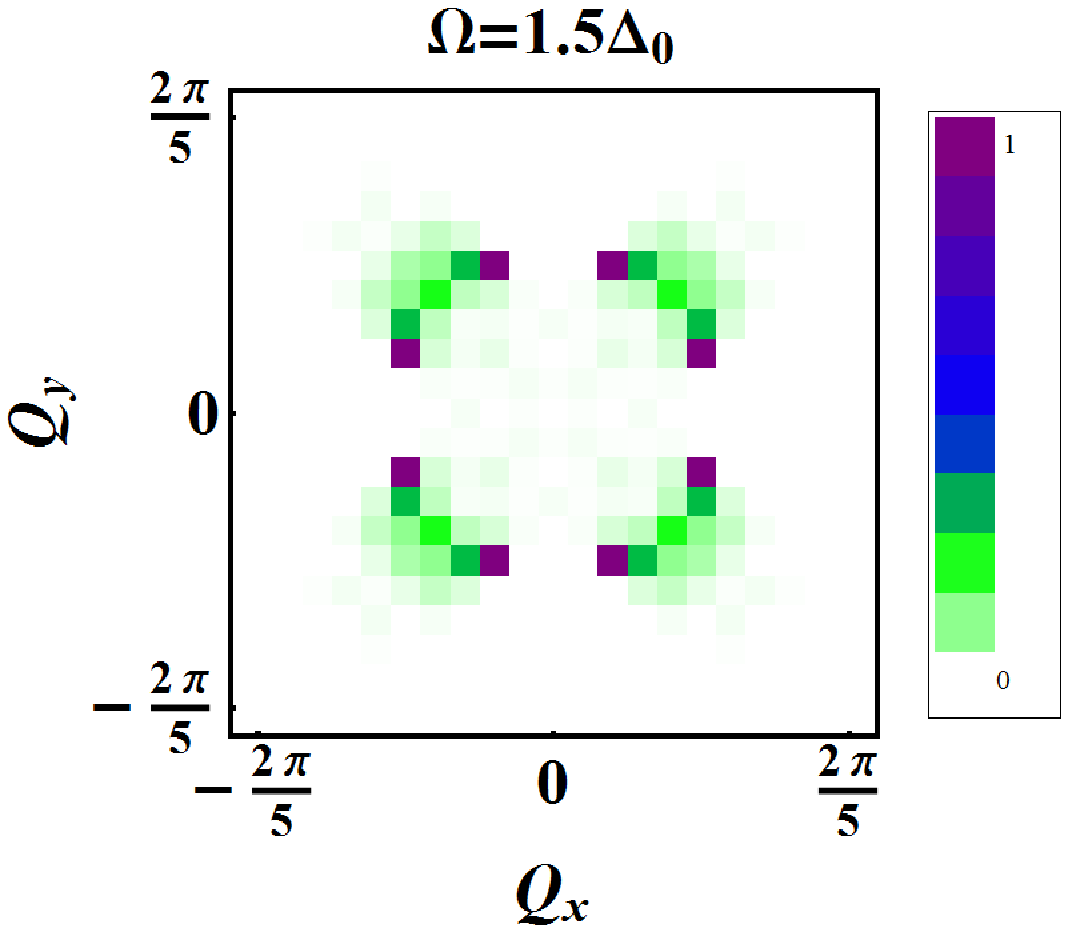} &
\includegraphics[width=0.31\textwidth,angle=0]{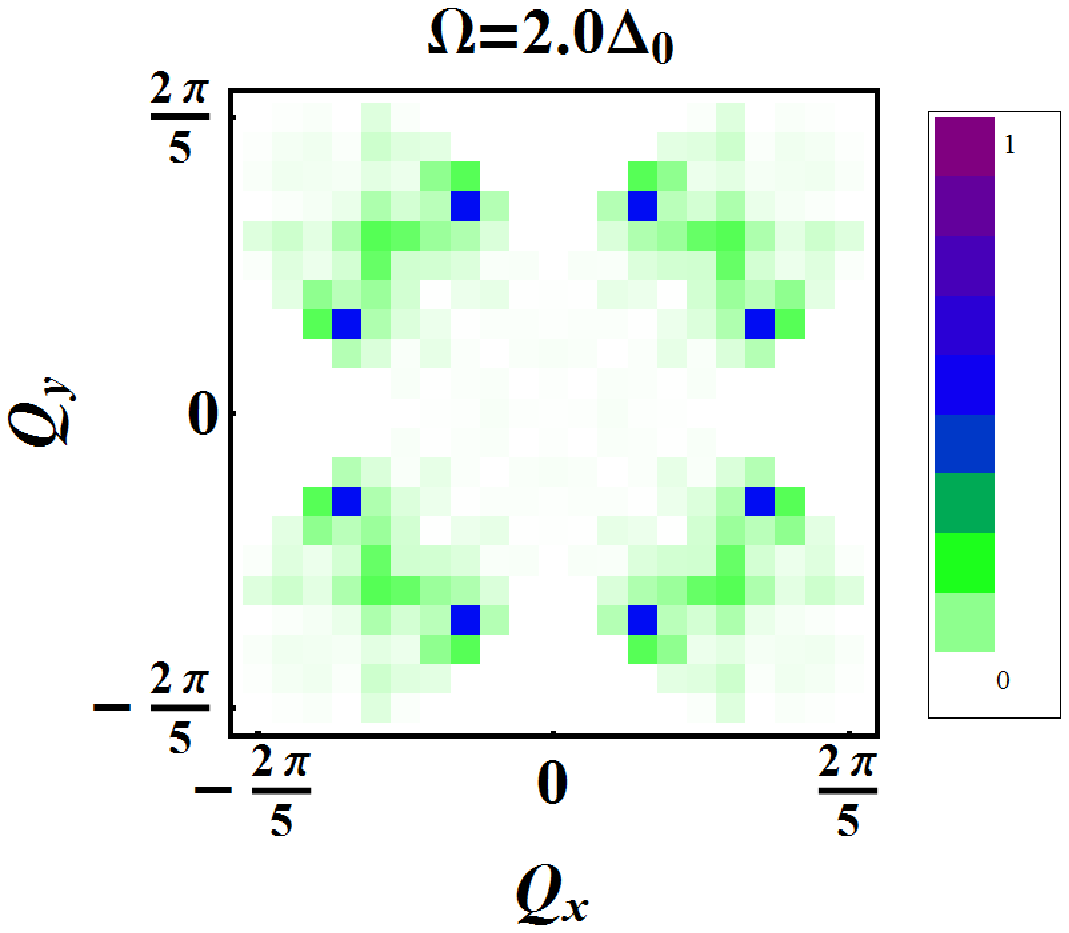}
\end{tabular}
\caption[Susceptibilities]{(Color Online) Plot of $Im[\Pi_{22}(0,0)-\Pi_{22}(\Q,\Omega)]^{-1}$ in the d-wave state  for (a) $\Omega=\Delta_0$ (b) $\Omega=1.5\Delta_0$ (c) $\Omega=2.0\Delta_0$. The poles in the denominator of the second term in Eq.~(\ref{RPAresp}) illustrate the Anderson-Bogoliubov mode. This result differs from the s-wave Anderson-Bogoliubov mode, Fig. \ref{NormalSwave}, due to the presence of a d-wave gap.
}
\label{CollMode}
\end{figure*}

The response functions calculated in this paper are complicated ``speckle" patterns in Q-space, shown in Figs.~\ref{SpinSusc} and \ref{densityResp}. Define the coherence factor $\eta_{ab}(\k,\Q)\equiv\left(1-a\frac{\xi_\k \xi_{\k+\Q}}{E_\k E_{\k+\Q}}-b\frac{\Delta_\k \Delta_{\k+\Q}}{E_\k E_{\k+\Q}}\right)$.
The response function can be interpreted from Eq.~\ref{Pi0ijab} as a convolution proportional to:
\[ \rm{Im} ~\Pi^{ab} \propto \int d\omega_1 d\omega_2 \sum_k \eta_{ab}(\k,\Q) \]
\begin{equation}
\times  \delta(\Omega-\omega_1-\omega_2)\delta(\omega_1-E_{\k})\delta(\omega_2-E_{\k+\Q}).
\label{DOSintegrals}
\end{equation}

\begin{figure}[tb]
\includegraphics[width=0.34\textwidth,angle=0]{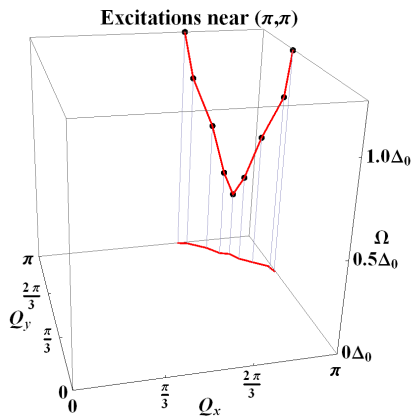} 
\caption[]{(Color Online) Positions of the maxima in $\rm{Im}\Pi_{33}$ around the point  $(\pi,\pi)$ in $\Q-\Omega$ space.}
\label{PiPi}
\end{figure}

The response function is a coherence-factor-weighted sum over the delta function. The single particle density of states (DOS) is $\rm{DOS}(\omega)=-\frac{1}{\pi}\sum_k \rm{Im} G(\k,\omega)=\sum_k \delta(\omega-E_{\k})$. If the frequency integrals in Eq.~\ref{DOSintegrals} are performed then we can also define a joint density of states (JDOS), for two particle excitations as
\begin{equation}
\rm{JDOS}(\Q,\Omega)=\sum_k  \delta(\Omega-E_{\k+\Q}-E_{\k}).
\label{JDOSequation}
\end{equation}
By examining the structure of the momentum resolved DOS and JDOS, meaning $\rm{DOS}(\k,\omega)\equiv\delta(\omega-E_{\k})$ for fixed $\omega$ without a sum over k, and using reasoning analogous to the octet model~\cite{Octet,Capriotti} from cuprate tunneling experiments~\cite{SeamusDavis}, the structure of the response functions becomes clear. We show plots of $\rm{DOS}(\k,\omega)$ for a d-wave superconductor defined by Eq.(\ref{green}) in Fig.~\ref{dwaveBZ}. The essential observation is that the final observed signal at wave vector $\Q$ will have a high transition probability when $\rm{DOS}(\k,\omega)=\rm{DOS}(\k+\Q,\omega+\Omega)$. So we compare the two densities of states. The available states in the $\rm{DOS}(\k,\omega)$ convolve to give rise to the JDOS (Fig.~\ref{JDOS}), which is then, in turn, weighted by a coherence factor in the dominant contribution to the response function. In tunneling experiments the scattering was elastic in contrast to this inelastic process transferring energy $\Omega$ from the state at $E_{\k}$ to the state at $E_{\k+\Q}$, so we compare $\rm{DOS}(\k,\omega)=\rm{DOS}(\k+\Q,\omega)$. This differs from the inelastic case which blurs the DOS over a range of allowed frequencies.

In Fig.~\ref{JDOS}, we have overlayed the JDOS for several fixed external frequencies with the response function. There is an overall qualitative agreement between the $\Q$ which nest the single $\rm{DOS}(\k,\omega)$ and $\rm{DOS}(\k+\Q,\omega+\Omega)$ and the JDOS's structure. The gross features of the JDOS (Fig.~\ref{JDOS}) are contained in the final response function (Figs.~\ref{SpinSusc} and ~\ref{densityResp}) which differ by a coherence factor. In both figures ~\ref{SpinSusc} and ~\ref{densityResp}, we see nodal excitations low $(\Omega,\Q)$ along the diagonals as well as transitions at finite $\Q$ resulting from the convolutions of the DOS.

The collective mode, shown in Fig.~\ref{CollMode} has a structure beyond the simple s-wave result due to it being d-wave. To check this, the poles in second term in Eq.~(\ref{RPAresp}), $\rm{Im}[\Pi_{22}(0,0)-\Pi_{22}(\Q,\Omega)]^{-1}$, can be plotted by itself in Fig.~\ref{CollMode}. These excitations are the d-wave equivalent of the Anderson-Bogoliubov mode. Unlike the s-wave Anderson-Bogoliubov mode where all quasiparticle excitations are gap, the d-wave Anderson-Bogoliubov mode is not isolated from the gapless quasiparticle response in the full $\rm{Im}[\Pi_{33}]$.

An additional observation can be made. In the inelastic neutron response for cuprates, excitations around $(\pi,\pi)$ have played an important role \cite{Zaanen}. We chose not to focus on the $(0,\pi)$ excitations which are broader and weaker than the $(\pi,\pi)$ excitations. Consequently, we are curious to see what comes out of the RPA analysis performed here. We have tracked the peak excitations which occur around the point $(\pi,\pi)$ for a limited range of frequencies. Within the limitations imposed by sampling discrete frequencies, we find that clear peak excitations around $(\pi,\pi)$ are absent for low energy, begin along the zone diagonal, and split into two peaks as we increase frequency. At the pair-breaking energy $2\Delta_0=0.4 t$ those excitations are near the Brillouin zone boundary, and above that energy they are difficult to identify amidst all the allowed excitations. This is shown in  Fig.~\ref{PiPi}.

We have shown by taking apart Eq.~\ref{DOSintegrals} that a combination of the geometry of the Fermi surface and the quasiparticle dispersion determines the ``speckle" patterns in Figs.~\ref{SpinSusc} and ~\ref{densityResp}. Notwithstanding the complicated nature of both the spin susceptibility and density-density response, the detailed k-space features should be specific enough to identify exotic pairing in superfluids distinctly from other states.

\section{Conclusion}

In summary, we have provided an analysis of two response functions, the spin susceptibility and the density-density response, for the inelastic scattering of light off a d-wave condensate. 

We included the effect of collective modes and examined their dispersion in the s-wave and d-wave cases. In the s-wave case, the collective mode is the only mode present below the pair breaking energy $2\Delta_0$, unencumbered by effects from the Coulomb interaction and disorder in the context of cold atom experiments. In the d-wave case, the collective mode is superimposed with other gapless excitations and has a structure beyond the simple s-wave case.

For the d-wave spin and density reponse, we provide an interpretation of the detailed momentum-space structure. The inelastic response functions reflect the opening of a d-wave gap on top of the geometry of the Fermi surface. These patterns are specific enough to unambiguously identify d-wave superfluidity, were it to arise, providing a cold-atom analog to phase-sensitive experiments in solids. We also examined the peak response near the point $\Q=(\pi,\pi)$ in analogy to the hourglass in cuprate neutron experiments \cite{Zaanen}. The primary result of our work is a precise tool, the momentum space structure of inelastic light scattering, to identify and characterize a d-wave paired Fermi superfluid which can be generalized to any gap symmetry.

\section{Acknowledgements}
GRB thanks Lev Bishop and P.~J.~Hirschfeld for useful discussions. GRB acknowledges support from the ARO's atomtronics MURI and VG acknowledges support from the US-ARO.

\bibliographystyle{apsrev}
\bibliography{braggBiblio}

\begin{thebibliography}{30}
\expandafter\ifx\csname natexlab\endcsname\relax\def\natexlab#1{#1}\fi
\expandafter\ifx\csname bibnamefont\endcsname\relax
  \def\bibnamefont#1{#1}\fi
\expandafter\ifx\csname bibfnamefont\endcsname\relax
  \def\bibfnamefont#1{#1}\fi
\expandafter\ifx\csname citenamefont\endcsname\relax
  \def\citenamefont#1{#1}\fi
\expandafter\ifx\csname url\endcsname\relax
  \def\url#1{\texttt{#1}}\fi
\expandafter\ifx\csname urlprefix\endcsname\relax\def\urlprefix{URL }\fi
\providecommand{\bibinfo}[2]{#2}
\providecommand{\eprint}[2][]{\url{#2}}

\bibitem[{\citenamefont{Bloch et~al.}(2008)\citenamefont{Bloch, Dalibard, and
  Zwerger}}]{IBloch}
\bibinfo{author}{\bibfnamefont{I.}~\bibnamefont{Bloch}},
  \bibinfo{author}{\bibfnamefont{J.}~\bibnamefont{Dalibard}}, \bibnamefont{and}
  \bibinfo{author}{\bibfnamefont{W.}~\bibnamefont{Zwerger}},
  \bibinfo{journal}{Rev. Mod. Phys.} \textbf{\bibinfo{volume}{80}},
  \bibinfo{pages}{885} (\bibinfo{year}{2008}).

\bibitem[{\citenamefont{Hofstetter et~al.}(2002)\citenamefont{Hofstetter,
  Cirac, Zoller, Demler, and Lukin}}]{Hofstetter}
\bibinfo{author}{\bibfnamefont{W.}~\bibnamefont{Hofstetter}},
  \bibinfo{author}{\bibfnamefont{J.~I.} \bibnamefont{Cirac}},
  \bibinfo{author}{\bibfnamefont{P.}~\bibnamefont{Zoller}},
  \bibinfo{author}{\bibfnamefont{E.}~\bibnamefont{Demler}}, \bibnamefont{and}
  \bibinfo{author}{\bibfnamefont{M.~D.} \bibnamefont{Lukin}},
  \bibinfo{journal}{Phys. Rev. Lett.} \textbf{\bibinfo{volume}{89}},
  \bibinfo{pages}{220407} (\bibinfo{year}{2002}).

\bibitem[{\citenamefont{Zhang et~al.}(1999)\citenamefont{Zhang, Sackett, and
  Hulet}}]{ZhangSackettHulet}
\bibinfo{author}{\bibfnamefont{W.}~\bibnamefont{Zhang}},
  \bibinfo{author}{\bibfnamefont{C.~A.} \bibnamefont{Sackett}},
  \bibnamefont{and} \bibinfo{author}{\bibfnamefont{R.~G.} \bibnamefont{Hulet}},
  \bibinfo{journal}{Phys. Rev. A} \textbf{\bibinfo{volume}{60}},
  \bibinfo{pages}{504} (\bibinfo{year}{1999}).

\bibitem[{\citenamefont{Combescot et~al.}(2006)\citenamefont{Combescot, Kagan,
  and Stringari}}]{Combescot}
\bibinfo{author}{\bibfnamefont{R.}~\bibnamefont{Combescot}},
  \bibinfo{author}{\bibfnamefont{M.~Y.} \bibnamefont{Kagan}}, \bibnamefont{and}
  \bibinfo{author}{\bibfnamefont{S.}~\bibnamefont{Stringari}},
  \bibinfo{journal}{Phys. Rev. A} \textbf{\bibinfo{volume}{74}},
  \bibinfo{pages}{042717} (\bibinfo{year}{2006}).

\bibitem[{\citenamefont{Minguzzi et~al.}(2001)\citenamefont{Minguzzi, Ferrari,
  and Castin}}]{Minguzzi}
\bibinfo{author}{\bibfnamefont{A.}~\bibnamefont{Minguzzi}},
  \bibinfo{author}{\bibfnamefont{G.}~\bibnamefont{Ferrari}}, \bibnamefont{and}
  \bibinfo{author}{\bibfnamefont{Y.}~\bibnamefont{Castin}},
  \bibinfo{journal}{The Euro.~Phys.~ J. D.} \textbf{\bibinfo{volume}{17}},
  \bibinfo{pages}{49} (\bibinfo{year}{2001}), ISSN \bibinfo{issn}{1434-6060}.

\bibitem[{\citenamefont{Buchler et~al.}(2004)\citenamefont{Buchler, Zoller, and
  Zwerger}}]{BuchlerZollerZwerger}
\bibinfo{author}{\bibfnamefont{H.~P.} \bibnamefont{Buchler}},
  \bibinfo{author}{\bibfnamefont{P.}~\bibnamefont{Zoller}}, \bibnamefont{and}
  \bibinfo{author}{\bibfnamefont{W.}~\bibnamefont{Zwerger}},
  \bibinfo{journal}{Phys. Rev. Lett.} \textbf{\bibinfo{volume}{93}},
  \bibinfo{pages}{080401} (\bibinfo{year}{2004}).

\bibitem[{\citenamefont{Ohashi}()}]{OhashiGriffin}
\bibinfo{author}{\bibfnamefont{Y.}~\bibnamefont{Ohashi}},
  \bibinfo{howpublished}{arXiv:cond-mat/0503641}.

\bibitem[{\citenamefont{Altman et~al.}(2004)\citenamefont{Altman, Demler, and
  Lukin}}]{AltmanDemlerLukin}
\bibinfo{author}{\bibfnamefont{E.}~\bibnamefont{Altman}},
  \bibinfo{author}{\bibfnamefont{E.}~\bibnamefont{Demler}}, \bibnamefont{and}
  \bibinfo{author}{\bibfnamefont{M.~D.} \bibnamefont{Lukin}},
  \bibinfo{journal}{Phys. Rev. A} \textbf{\bibinfo{volume}{70}},
  \bibinfo{pages}{013603} (\bibinfo{year}{2004}).

\bibitem[{\citenamefont{Pekker}()}]{Pekker}
\bibinfo{author}{\bibfnamefont{D.}~\bibnamefont{Pekker}},
  \bibinfo{howpublished}{arXiv:0906.0931v1}.

\bibitem[{\citenamefont{Lutchyn et~al.}(2008)\citenamefont{Lutchyn, Nagornykh,
  and Yakovenko}}]{Yakovenko}
\bibinfo{author}{\bibfnamefont{R.~M.} \bibnamefont{Lutchyn}},
  \bibinfo{author}{\bibfnamefont{P.}~\bibnamefont{Nagornykh}},
  \bibnamefont{and} \bibinfo{author}{\bibfnamefont{V.~M.}
  \bibnamefont{Yakovenko}}, \bibinfo{journal}{Phys. Rev. B}
  \textbf{\bibinfo{volume}{77}}, \bibinfo{pages}{144516}
  (\bibinfo{year}{2008}).

\bibitem[{\citenamefont{Pitaevskii}(2003)}]{BECbook}
\bibinfo{author}{\bibfnamefont{L.~P.} \bibnamefont{Pitaevskii}},
  \emph{\bibinfo{title}{Bose-Einstein Condensation}}
  (\bibinfo{publisher}{Oxford University Press}, \bibinfo{year}{2003}).

\bibitem[{\citenamefont{Stamper-Kurn et~al.}(1999)\citenamefont{Stamper-Kurn,
  Chikkatur, G\"orlitz, Inouye, Gupta, Pritchard, and Ketterle}}]{Stamper-Kurn}
\bibinfo{author}{\bibfnamefont{D.~M.} \bibnamefont{Stamper-Kurn}},
  \bibinfo{author}{\bibfnamefont{A.~P.} \bibnamefont{Chikkatur}},
  \bibinfo{author}{\bibfnamefont{A.}~\bibnamefont{G\"orlitz}},
  \bibinfo{author}{\bibfnamefont{S.}~\bibnamefont{Inouye}},
  \bibinfo{author}{\bibfnamefont{S.}~\bibnamefont{Gupta}},
  \bibinfo{author}{\bibfnamefont{D.~E.} \bibnamefont{Pritchard}},
  \bibnamefont{and} \bibinfo{author}{\bibfnamefont{W.}~\bibnamefont{Ketterle}},
  \bibinfo{journal}{Phys. Rev. Lett.} \textbf{\bibinfo{volume}{83}},
  \bibinfo{pages}{2876} (\bibinfo{year}{1999}).

\bibitem[{\citenamefont{Corcovilos et~al.}(2010)\citenamefont{Corcovilos, Baur,
  Hitchcock, Mueller, and Hulet}}]{HuletPaper}
\bibinfo{author}{\bibfnamefont{T.~A.} \bibnamefont{Corcovilos}},
  \bibinfo{author}{\bibfnamefont{S.~K.} \bibnamefont{Baur}},
  \bibinfo{author}{\bibfnamefont{J.~M.} \bibnamefont{Hitchcock}},
  \bibinfo{author}{\bibfnamefont{E.~J.} \bibnamefont{Mueller}},
  \bibnamefont{and} \bibinfo{author}{\bibfnamefont{R.~G.} \bibnamefont{Hulet}},
  \bibinfo{journal}{Phys. Rev. A} \textbf{\bibinfo{volume}{81}},
  \bibinfo{pages}{013415} (\bibinfo{year}{2010}).

\bibitem[{\citenamefont{Brunello et~al.}(2001)\citenamefont{Brunello, Dalfovo,
  Pitaevskii, Stringari, and Zambelli}}]{MomentumBragg}
\bibinfo{author}{\bibfnamefont{A.}~\bibnamefont{Brunello}},
  \bibinfo{author}{\bibfnamefont{F.}~\bibnamefont{Dalfovo}},
  \bibinfo{author}{\bibfnamefont{L.}~\bibnamefont{Pitaevskii}},
  \bibinfo{author}{\bibfnamefont{S.}~\bibnamefont{Stringari}},
  \bibnamefont{and} \bibinfo{author}{\bibfnamefont{F.}~\bibnamefont{Zambelli}},
  \bibinfo{journal}{Phys. Rev. A} \textbf{\bibinfo{volume}{64}},
  \bibinfo{pages}{063614} (\bibinfo{year}{2001}).

\bibitem[{\citenamefont{Kulik et~al.}(1981)\citenamefont{Kulik, Entin-Wohlman,
  and Orbach}}]{KEO}
\bibinfo{author}{\bibfnamefont{I.~O.} \bibnamefont{Kulik}},
  \bibinfo{author}{\bibfnamefont{O.}~\bibnamefont{Entin-Wohlman}},
  \bibnamefont{and} \bibinfo{author}{\bibfnamefont{R.}~\bibnamefont{Orbach}},
  \bibinfo{journal}{J. Low Temp. Phys.} \textbf{\bibinfo{volume}{43}},
  \bibinfo{pages}{591} (\bibinfo{year}{1981}).

\bibitem[{\citenamefont{Wong and Takada}(1988)}]{WongTakada}
\bibinfo{author}{\bibfnamefont{K.~Y.~M.} \bibnamefont{Wong}} \bibnamefont{and}
  \bibinfo{author}{\bibfnamefont{S.}~\bibnamefont{Takada}},
  \bibinfo{journal}{Phys. Rev. B} \textbf{\bibinfo{volume}{37}},
  \bibinfo{pages}{5644} (\bibinfo{year}{1988}).

\bibitem[{\citenamefont{Ohashi and Takada}(2000)}]{OhashiTakada}
\bibinfo{author}{\bibfnamefont{Y.}~\bibnamefont{Ohashi}} \bibnamefont{and}
  \bibinfo{author}{\bibfnamefont{S.}~\bibnamefont{Takada}},
  \bibinfo{journal}{Phys. Rev. B} \textbf{\bibinfo{volume}{62}},
  \bibinfo{pages}{5971} (\bibinfo{year}{2000}).

\bibitem[{\citenamefont{Ohashi and Takada}(1997)}]{OhashiTakada2}
\bibinfo{author}{\bibfnamefont{Y.}~\bibnamefont{Ohashi}} \bibnamefont{and}
  \bibinfo{author}{\bibfnamefont{S.}~\bibnamefont{Takada}},
  \bibinfo{journal}{J.~Phys.~ Soc.~Jpn.} \textbf{\bibinfo{volume}{66}},
  \bibinfo{pages}{2437} (\bibinfo{year}{1997}).

\bibitem[{\citenamefont{Zou et~al.}(2010)\citenamefont{Zou, Kuhnle, Vale, and
  Hu}}]{BraggZouHuiHu}
\bibinfo{author}{\bibfnamefont{P.}~\bibnamefont{Zou}},
  \bibinfo{author}{\bibfnamefont{E.~D.} \bibnamefont{Kuhnle}},
  \bibinfo{author}{\bibfnamefont{C.~J.} \bibnamefont{Vale}}, \bibnamefont{and}
  \bibinfo{author}{\bibfnamefont{H.}~\bibnamefont{Hu}}, \bibinfo{journal}{Phys.
  Rev. A} \textbf{\bibinfo{volume}{82}}, \bibinfo{pages}{061605}
  (\bibinfo{year}{2010}).

\bibitem[{\citenamefont{Mazin and Yakovenko}(1995)}]{MazinYakovenko}
\bibinfo{author}{\bibfnamefont{I.~I.} \bibnamefont{Mazin}} \bibnamefont{and}
  \bibinfo{author}{\bibfnamefont{V.~M.} \bibnamefont{Yakovenko}},
  \bibinfo{journal}{Phys. Rev. Lett.} \textbf{\bibinfo{volume}{75}},
  \bibinfo{pages}{4134} (\bibinfo{year}{1995}).

\bibitem[{\citenamefont{Lu}(1992)}]{PhysRevLett.68.125}
\bibinfo{author}{\bibfnamefont{J.~P.} \bibnamefont{Lu}},
  \bibinfo{journal}{Phys. Rev. Lett.} \textbf{\bibinfo{volume}{68}},
  \bibinfo{pages}{125} (\bibinfo{year}{1992}).

\bibitem[{\citenamefont{Kao et~al.}(2000)\citenamefont{Kao, Si, and
  Levin}}]{KLevinNeutron}
\bibinfo{author}{\bibfnamefont{Y.-J.} \bibnamefont{Kao}},
  \bibinfo{author}{\bibfnamefont{Q.}~\bibnamefont{Si}}, \bibnamefont{and}
  \bibinfo{author}{\bibfnamefont{K.}~\bibnamefont{Levin}},
  \bibinfo{journal}{Phys. Rev. B} \textbf{\bibinfo{volume}{61}},
  \bibinfo{pages}{R11898} (\bibinfo{year}{2000}).

\bibitem[{\citenamefont{Lavagna and Stemmann}(1994)}]{PhysRevB.49.4235}
\bibinfo{author}{\bibfnamefont{M.}~\bibnamefont{Lavagna}} \bibnamefont{and}
  \bibinfo{author}{\bibfnamefont{G.}~\bibnamefont{Stemmann}},
  \bibinfo{journal}{Phys. Rev. B} \textbf{\bibinfo{volume}{49}},
  \bibinfo{pages}{4235} (\bibinfo{year}{1994}).

\bibitem[{\citenamefont{Liu et~al.}(1995)\citenamefont{Liu, Zha, and
  Levin}}]{PhysRevLett.75.4130}
\bibinfo{author}{\bibfnamefont{D.~Z.} \bibnamefont{Liu}},
  \bibinfo{author}{\bibfnamefont{Y.}~\bibnamefont{Zha}}, \bibnamefont{and}
  \bibinfo{author}{\bibfnamefont{K.}~\bibnamefont{Levin}},
  \bibinfo{journal}{Phys. Rev. Lett.} \textbf{\bibinfo{volume}{75}},
  \bibinfo{pages}{4130} (\bibinfo{year}{1995}).

\bibitem[{\citenamefont{Zha et~al.}(1993)\citenamefont{Zha, Levin, and
  Si}}]{PhysRevB.47.9124}
\bibinfo{author}{\bibfnamefont{Y.}~\bibnamefont{Zha}},
  \bibinfo{author}{\bibfnamefont{K.}~\bibnamefont{Levin}}, \bibnamefont{and}
  \bibinfo{author}{\bibfnamefont{Q.}~\bibnamefont{Si}}, \bibinfo{journal}{Phys.
  Rev. B} \textbf{\bibinfo{volume}{47}}, \bibinfo{pages}{9124}
  (\bibinfo{year}{1993}).

\bibitem[{\citenamefont{Belkhir and Randeria}(1994)}]{RanderiaABdispersion}
\bibinfo{author}{\bibfnamefont{L.}~\bibnamefont{Belkhir}} \bibnamefont{and}
  \bibinfo{author}{\bibfnamefont{M.}~\bibnamefont{Randeria}},
  \bibinfo{journal}{Phys. Rev. B} \textbf{\bibinfo{volume}{49}},
  \bibinfo{pages}{6829} (\bibinfo{year}{1994}).

\bibitem[{\citenamefont{Wang and Lee}(2003)}]{Octet}
\bibinfo{author}{\bibfnamefont{Q.-H.} \bibnamefont{Wang}} \bibnamefont{and}
  \bibinfo{author}{\bibfnamefont{D.-H.} \bibnamefont{Lee}},
  \bibinfo{journal}{Phys. Rev. B} \textbf{\bibinfo{volume}{67}},
  \bibinfo{pages}{020511} (\bibinfo{year}{2003}).

\bibitem[{\citenamefont{Capriotti et~al.}(2003)\citenamefont{Capriotti,
  Scalapino, and Sedgewick}}]{Capriotti}
\bibinfo{author}{\bibfnamefont{L.}~\bibnamefont{Capriotti}},
  \bibinfo{author}{\bibfnamefont{D.~J.} \bibnamefont{Scalapino}},
  \bibnamefont{and} \bibinfo{author}{\bibfnamefont{R.~D.}
  \bibnamefont{Sedgewick}}, \bibinfo{journal}{Phys. Rev. B}
  \textbf{\bibinfo{volume}{68}}, \bibinfo{pages}{014508}
  (\bibinfo{year}{2003}).

\bibitem[{\citenamefont{Hoffman et~al.}(2002)\citenamefont{Hoffman, McElroy,
  Lee, Lang, Eisaki, Uchida, and Davis}}]{SeamusDavis}
\bibinfo{author}{\bibfnamefont{J.~E.} \bibnamefont{Hoffman}},
  \bibinfo{author}{\bibfnamefont{K.}~\bibnamefont{McElroy}},
  \bibinfo{author}{\bibfnamefont{D.-H.} \bibnamefont{Lee}},
  \bibinfo{author}{\bibfnamefont{K.~M.} \bibnamefont{Lang}},
  \bibinfo{author}{\bibfnamefont{H.}~\bibnamefont{Eisaki}},
  \bibinfo{author}{\bibfnamefont{S.}~\bibnamefont{Uchida}}, \bibnamefont{and}
  \bibinfo{author}{\bibfnamefont{J.~C.} \bibnamefont{Davis}},
  \bibinfo{journal}{Science} \textbf{\bibinfo{volume}{297}},
  \bibinfo{pages}{1148} (\bibinfo{year}{2002}).

\bibitem[{\citenamefont{Zaanen}(2011)}]{Zaanen}
\bibinfo{author}{\bibfnamefont{J.}~\bibnamefont{Zaanen}},
  \bibinfo{journal}{Nature} \textbf{\bibinfo{volume}{471}},
  \bibinfo{pages}{314} (\bibinfo{year}{2011}).

\end{thebibliography}

\end{document}